\documentclass[12pt]{iopart}
\usepackage{cite}
\usepackage[usenames,dvipsnames]{xcolor}
\definecolor{ao(english)}{rgb}{0.0, 0.5, 0.0}
\usepackage[breaklinks,colorlinks=true,linkcolor=blue,citecolor=ao(english)]{hyperref}
\usepackage[title,titletoc]{appendix}
\usepackage[english]{babel}
\usepackage[utf8]{inputenc}
\usepackage[OT2,T1]{fontenc}
\usepackage{graphicx}
\usepackage{amssymb,amsfonts,amsthm}
\usepackage{iopams}
\usepackage{bm,bbm}
\usepackage{tikz,stackrel}
\usepackage{pstricks}
\usepackage{pifont} 
\usepackage{changepage}
\usepackage{silence} 
\newcommand{\bc}{\begin{center}}
\newcommand{\ec}{\end{center}}
\newcommand{\be}{\begin{equation}}
\newcommand{\ee}{\end{equation}}
\newcommand{\ba}{\begin{eqnarray}}
\newcommand{\ea}{\end{eqnarray}}
\def\bs{\begin{subequations}}
\def\es{\end{subequations}}
\renewcommand{\leq}{\leqslant}
\renewcommand{\geq}{\geqslant}
\def\a{\alpha}
\def\b{\beta}
\def\de{\delta}
\def\g{\gamma}
\def\la{\lambda}
\def\k{\kappa}
\def\e{\epsilon}
\def\ve{\varepsilon}
\def\Om{\Omega}
\def\om{\omega}
\def\De{\Delta}
\def\G{\Gamma}
\def\t{\tau}  
\def\s{\sigma}

\def\vp{\varphi}

\def\cA{\mathcal{A}}
\def\cB{\mathcal{B}}
\def\cC{\mathcal{C}}
\def\cD{\mathcal{D}}

\def\cF{\mathcal{F}}

\def\cJ{\mathcal{J}}
\def\cK{\mathcal{K}}
\def\cL{\mathcal{L}}

\def\cV{\mathcal{V}}

\def\ds{d_\textsc{s}}
\def\dh{d_\textsc{h}}

\def\p{\partial}
\def\bp{\bar{\partial}}
\def\B{\Box}
\newcommand{\Eq}[1]{(\ref{#1})}

\def\cob{\color{blue}}
\newcommand{\au}[2]{#2 #1}
\newcommand{\ua}[2]{#1 #2}
\newcommand{\book}[5]{\emph{#1} (#3: #2)}
\newcommand{\books}[4]{\emph{#1} (#3: #2)}
\newcommand{\oarX}[1]{\href{http://arxiv.org/abs/#1}{{\ttfamily\cob arXiv:#1}}}
\newcommand{\arX}[1]{\href{http://arxiv.org/abs/#1}{{\ttfamily\cob arXiv:#1}}}
\newcommand{\doin}[6]{\href{http://dx.doi.org/#1}{{\cob {\it #2} #3 {\bf #4} #5}}}
\newcommand{\doinn}[5]{\href{http://dx.doi.org/#1}{{\cob {\it #2} {\bf #3} #4}}}
\newcommand{\doij}[5]{\href{http://dx.doi.org/#1}{{\cob {\it #2} #3(#5)#4}}}
\newcommand{\ndoin}[6]{\href{#1}{{\cob {\it #2} #3 {\bf #4} #5}}}
\newcommand{\ndoinn}[5]{\href{#1}{{\cob {\it #2} {\bf #3} #4}}}
\newcommand{\procsinm}[5]{\emph{#1} ed #2 (#4: #3)}

\newcommand{\tia}[1]{#1}
\newcommand{\boxd}[1]{\fbox{$\displaystyle\phantom{\Biggl(}#1\phantom{\Biggl)}$}}

\def\lp{\ell_{\rm Pl}}

\def\ep{E_{\rm Pl}}
\def\Re{{\rm Re}}
\def\Im{{\rm Im}}

\newcounter{listcounter}

\makeatletter
\long\def\@makefntext#1{\parindent 1em\noindent 
 \makebox[1em][l]{\footnotesize\rm$\m@th{^\arabic{footnote}}$}%
 \footnotesize\rm #1}
\def\@makefnmark{\hbox{$^{\arabic{footnote}}\m@th$}}
\def\@thefnmark{\arabic{footnote}}
\makeatother

\begin{document}

\title{Quantum scalar field theories with fractional operators}

\author{Gianluca Calcagni}
\address{Instituto de Estructura de la Materia, CSIC, Serrano 121, 28006 Madrid, Spain}
\ead{g.calcagni@csic.es}
\vspace{10pt}
\begin{indented}
\item[]February 3, 2021
\end{indented}


\begin{abstract}
We study a class of perturbative scalar quantum field theories where dynamics is characterized by Lorentz-invariant or Lorentz-breaking non-local operators of fractional order and the underlying spacetime has a varying spectral dimension. These theories are either ghost free or power-counting renormalizable but they cannot be both at the same time. However, some of them are one-loop unitary and finite, and possibly unitary and finite at all orders.
\end{abstract}




\centerline{2021 \doinn{10.1088/1361-6382/ac103c}{Class.\ Quantum Grav.}{38}{165006}{2021} [\arX{2102.03363}]}


\tableofcontents


\section{Introduction}\label{intro}

Some time ago, it was noted that all theories of quantum gravity are characterized by dimensional flow: whenever a continuous spacetime arises, fundamentally or effectively, its dimension varies with the probed scale \cite{tH93,Car09,fra1}. Since a varying dimensionality requires spacetime geometry to be endowed with intrinsic time and length scales, a theory with dimensional flow is said to be \emph{multi-scale}. All quantum gravities are multi-scale in certain corners of their parameter space and many examples of dimensional flow in quantum gravity have been studied, or re-studied, since then \cite{trtls,revmu,Car17,MiTr}. Among other features, it emerged that not all multi-scale spacetimes can be called \emph{multi-fractal} indiscriminately, but only those whose Hausdorff, spectral and walk dimensions are related to one another in a certain way \cite{trtls}.

In parallel, it was suspected that dimensional flow could be related to the taming of infinities in the ultraviolet (UV). This possibility is natural in the view of the widely successful dimensional regularization scheme \cite{BG72,tV72}, according to which the divergence of four-dimensional integrals appearing in the summation of Feynman diagrams can be singled-out, and eventually subtracted by counter-terms, by artificially letting the topological dimension $D$ to vary, then setting it to $D=4$ at the end of the calculation after subtraction. If one could devise an exotic geometry with a physically meaningful non-integer dimension, the infinities of a quantum field theory (QFT) might be cured. This idea was investigated both in a model-independent fashion and in the specific framework of \emph{multi-fractional} theories \cite{revmu}, field theories living in multi-scale spacetimes whose measure and kinetic operators are factorizable in the coordinates. At first, having quantization of gravity in mind, the expectation was that a spacetime with reduced dimensionality in the UV would improve the renormalization properties of a QFT \cite{fra1}. However, two counter-examples were found where QFTs living in spacetimes with varying dimension\footnote{These examples are the multi-fractional theories dubbed $T_v$ and $T_q$ with, respectively, weighted and $q$-derivatives.
The first ($T_v$) lives in a spacetime with varying Hausdorff dimension and constant spectral dimension \cite{revmu}; this spacetime is not a multi-fractal. The second ($T_q$) lives in a multi-fractal spacetime with varying Hausdorff and spectral dimension \cite{frc7,revmu}.} were no better behaved than their ordinary analogue on standard Minkowski spacetime \cite{frc9}.

Since then, the trajectory of multi-fractional theories has been oscillating between phenomenology and the striving for their original goal of renormalizability. To understand what led us to the present point, a short historical digression may be helpful. Interest on fractal spacetimes ran for some time as a subject \emph{per se} \cite{Sti77,Ord83,NSc,ZS,ScM,MuS,Svo87,Ey89a,Ey89b,Not93,Not97}. Some proposals were inspired by dimensional regularization \cite{Sti77,ScM,MuS}, others \cite{Svo87,Ey89a,Ey89b} by early results about a quantum-gravity fractal foam \cite{CrSm1,CrSm2}, yet others attempted to reconcile quantum mechanics and general relativity by extending the principles of special relativity to intrinsic scales \cite{Not93,Not97}, while a few more were simply born out of pure curiosity \cite{Ord83,NSc,ZS}. Soon, however, attention was drawn to the booming and blooming of quantum-gravity theories and the discovery, between the early 1990s and the late 2000s, of their fractal-like properties at short scales \cite{ADJ,KKSW,KKMW,AmJ,AJW2,AJL4,LaR5,DJW1,Ben08,Mod08,Hor3,BeH,AA} (more references can be found in \cite{trtls,revmu,Car17,MiTr,frc2,mf2}). It was at this point that a general pattern in dimensional flow was recognized \cite{tH93,Car09} and research on field theories on spacetimes with varying dimension was revived from different perspectives involving fractional integration measures or fractional operators: QFTs on multi-fractal spacetimes \cite{fra1,fra2,fra3} were followed by 
 multi-fractional spacetimes \cite{frc2,fra4,frc1,first,CaRo2} (see \cite{revmu} for a review and more references) and, soon afterwards, Trinchero's model of scalar QFT \cite{Tri12,Tri17,Tri18} and a scalar model related to causal sets \cite{Belenchia:2014fda,Saravani:2015rva,Belenchia:2016sym,Belenchia:2017yvv,Saravani:2018rwm}. These models are indebted to the pioneering works on fractional powers of the d'Alembertian \cite{BGG,Mar91,Gia91,BGO,BG,doA92,Mar93,BOR,BBOR1,Barci:1996ny,BBOR2}. The circle closes because Bollini and Giambiagi, authors that studied the fractional d'Alembertian \cite{BGG,Gia91,BGO,BG}, also proposed dimensional regularization in the first place \cite{BG72}, in both cases with the objective of controlling the UV divergences appearing in QFT. 

Coming back to the multi-fractional paradigm, while the spacetime measure is uniquely defined parametrically \cite{revmu,first}, different choices of Lagrangian symmetries can produce three types of kinetic terms, either with ordinary but measure-dependent derivatives (theories $T_1$, $T_v$ and $T_q$, in the labeling of \cite{revmu}), or with fractional operators (collectively labeled $T_\g$), i.e., integro-differential operators studied in the branch of mathematics known as fractional calculus \cite{MR,Pod99,SKM,KST}. Although the first multi-fractional theories that were considered did employ fractional derivatives \cite{frc1,frc2,fra6,frc4}, later developments focused on the proposals $T_1$ with ordinary derivatives and $T_v$ and $T_q$ with measure-decorated ordinary derivatives, for the reason that they were more manageable and generated a wealth of easily testable phenomenology \cite{revmu}. After near completion of this program, we have recently turned our interest back to the original problem, left open in \cite{frc1,frc4}, of how to define a field theory with scale-dependent fractional derivatives \cite{revmu,mf0}, with the hope that it would not suffer from the non-renormalizability problems of the other multi-fractional candidates.

In this paper, we set the foundations of the class of multi-fractional theories $T_\g$ with fractional operators and work out some of their classical and quantum properties in the case of a scalar field. We find that there are six inequivalent theories in this class. A first sub-division depends on whether the kinetic operator is made of fractional derivatives or is the fractional Laplace--Beltrami (or d'Alembertian) operator. The first option violates Lorentz invariance, while the second preserves it. Then, for each case, one starts with a kinetic term made of just one multiplicative operator with fixed fractional exponent $\g$ (theories we call $T[\p^\g]$ and $T[\B^\g]$). Taken as stand-alone theories, $\g$ must be close to 1 in order to respect all known experimental constraints on Standard-Model and gravitational physical observables. However, $T[\p^\g]$ and $T[\B^\g]$ can also be taken as the basis to develop multi-fractional theories, either by adding ordinary derivative operators (theories $T[\p+\p^\g]$ and $T[\B+\B^\g]$) or by taking a scale-dependent fractional exponent (theories $T[\p^{\g(\ell)}]$ and $T[\B^{\g(\ell)}]$).

While $T[\p^\g]$ and $T[\B^\g]$ have constant Hausdorff and spectral dimension, the dimensional flow of all the other theories is the same, with a constant Hausdorff dimension and a varying spectral dimension. We will carry out a systematic study in the case of a real scalar field theory, which is a good training ground in preparation for gravity, which is touched upon in a companion paper \cite{mf2}. 

For the theory $T[\p^\g]$, we will show that the propagator in Lorentzian signature is non-analytic in momentum space, due to the presence of the absolute value $|k|$. To overcome this problem, one can take two roads: either one defines the quantum theory with complex-valued internal momenta $k^0$ and real-valued external momenta $p^0$ and integrates on open paths in the complex plane (Efimov analytic continuation \cite{Efimov:1967dpd,Pius:2016jsl,Briscese:2018oyx}) or, in alternative, one employs the traditional closed paths (contours) of local QFT but restricts the anomalous differential structure to spatial directions only, thus providing a sort of fractional extension of theories with higher-order spatial Laplacians, such as Ho\v{r}ava--Lifshitz gravity \cite{Hor09}. The multi-fractional versions $T[\p+\p^\g]$ and $T[\p^{\g(\ell)}]$ will also be sketched.

At the same time, we will propose an alternative (theory $T[\B^\g]$) where the propagator is analytic, Lorentz invariance is preserved, the dynamics is simpler and soluble and there are no ghosts for an order $\g\leq 1$ of the fractional operators. The theory is power-counting renormalizable only in the different range $\g>2$, thus complying with the original expectation of an improved renormalizability but at the unforeseen price of losing unitarity. Almost the converse holds for the other multi-fractional theories $T_v$ and $T_q$: they are always unitary\footnote{It can be easily checked with the method of appendix \ref{appB} using the momentum transform for these theories \cite{frc3,frc11}. This result amends the non-unitarity claim made in \cite{revmu}. We were unable to prove unitarity for the theory $T_1$.} but their power-counting renormalizability is never improved, at least for spacetime measures without logarithmic oscillations \cite{revmu}. Still, by explicit calculations we show that the theory $T[\B^\g]$ is one-loop finite and possibly finite at all orders in the range where it is also unitary, except for some specific values of $\g$. This implies that the theory $T[\B+\B^\g]$ cannot be unitary and renormalizable at the same time, a problem which could be solved by a realization of the multi-fractional \emph{Ansatz} via a scale-dependent operator (theory $T[\B^{\g(\ell)}]$).

The plan of the paper is as follows. In section \ref{sec2}, the general setting is introduced. The scalar theory $T[\p^\g]$ with fractional derivatives $\cD^\g$ is discussed in section \ref{sec3}, while the Lorentz-invariant scalar theory $T[\B^\g]$ with fractional d'Alembertian $(m^2-\B)^\g$, with or without mass, will be presented and analyzed in section \ref{sec4}. Sections \ref{sec3} and \ref{sec4} are independent and the reader unfamiliar with fractional calculus may skip section \ref{sec3} at the first reading and concentrate on $T[\B^\g]$, which is simpler and more intuitive than $T[\p^\g]$. A comparison with other results in the literature of scalar fields with fractional operators is made in section \ref{actimm}. Section \ref{conc} is devoted to conclusions. Appendix \ref{appA} contains technicalities about fractional derivatives, while in appendix \ref{appB} we give an alternative proof of unitarity of the Lorentz-invariant scalar theory $T[\B^\g]$.


\section{Multi-scale dynamics}\label{sec2}

In this section, we will introduce the main structure of the theory of quantum gravity we want to build. As a first step, we consider a scalar field theory and customize it to reproduce (i) an effect found in all quantum-gravity theories, namely, dimensional flow and (ii) a characteristic desired in a perturbative QFT approach, namely, power-counting renormalizability.

Consider a real scalar field $\phi$ living in $D$-dimensional Minkowski spacetime with metric $\eta_{\mu\nu}={\rm diag}(-,+,\cdots,+)$. An ordinary field theory with self-interaction $V(\phi)$ is governed by the action
\be\label{ordi}
S=\int\rmd^Dx\,\left[\frac12\phi\B\phi-V(\phi)\right],
\ee
where $\rmd^Dx$ is the Lebesgue measure in $D$ topological dimensions and $\B=\eta^{\mu\nu}\p_\mu\p_\nu$ is the d'Alembertian. In general, quantum-gravity effects can modify the integro-differential structure of \Eq{ordi} in the UV, namely, the measure $\rmd^Dx$ and the kinetic operator $\B$. These effects gradually disappear as the observer moves to IR lengths or low energy scales. The magnitude and nature of the modifications strongly depend on the specific theory and there may even be cases where geometry is not continuous at small scales and all corrections are negligible when the continuum limit \Eq{ordi} is reached. Here we will focus our attention on theories defined on a continuum and where the d'Alembertian is corrected by a fractional derivative term,
\be\label{kei}
\B\to \cK\simeq \ell_*^{2-2\g}\B+O(\p^{2\g})\,,
\ee
where $\ell_*$ is a fundamental length scale and $\g$ is a real constant. The ensuing theories can also be regarded as models of other quantum gravities in their continuum limit but here we will treat them as stand-alone proposals, thus demanding self-consistency in the form of unitarity and renormalizability.

With the aim of introducing the theory as a modification of standard QFT or gravity, it is more natural to attach the $\ell_*$ factor to the fractional-derivative term: $\cK\simeq \B+\ell_*^{2\g-2}O(\p^{2\g})$. However, for estimating UV divergences \Eq{kei} is slightly more convenient. The two choices are physically equivalent.


\subsection{Action}

The most general action for a real scalar with operators of fractional order in Minkowski spacetime is
\be\label{actss}
S=\int\rmd^Dx\,v(x)\left[\frac12\phi\cK\phi-V(\phi)\right],
\ee
where $v(x)$ is a non-trivial measure weight and $\cK$ is the kinetic operator \Eq{kei}.


\subsection{Choice of spacetime measure}

The measure weight $v(x)$ can be either 1 or a specific multi-scaling parametric profile dictated by basic dynamics-independent requirements on the Hausdorff dimension of spacetime \cite{revmu,first}. This profile is universal in the sense that it applies to any theory of quantum gravity, although each specific dynamics will fix the parameters in $v$ in a unique way. Two of these parameters are the UV scaling of the measure in the time direction ($\a_0\in\mathbb{R}$) and the UV scaling in the spatial directions ($\a\in\mathbb{R}$), so that
\be\label{scav}
[\rmd^Dx\,v]_\textsc{ir}=-D\,,\qquad [\rmd^Dx\,v]_\textsc{uv}=-\a_0-(D-1)\a\,.
\ee
where square brackets denote the engineering (energy) dimension. Throughout the paper, energy-momentum scales normally as $[k^\mu]=1$.

At the classical level, a non-trivial $v(x)$ does not present any major difficulty, but at the quantum level it leads to non-delta-like distributions for vertex contributions, which make the Feynman expansion challenging if not impossible \cite{frc9}. In some cases with simple kinetic terms, as the multi-fractional theories $T_v$ and $T_q$, this problem is avoided by a mathematical trick (a mapping to a non-physical frame where the QFT simplifies), but no such a stratagem exists for more complicated $\cK$s. Therefore, in this paper
\be\label{v1}
v(x)=1\,,\qquad \a_0=1=\a\,,
\ee
and the Hausdorff dimension of spacetime is constant, $\dh=-[\rmd^Dx\,v]=D$. In the rest of the section, however, for the sake of generality we will keep the scaling of the measure weight arbitrary, letting only $\a_0=\a$, so that $\dh=D\a$ in the UV.


\subsection{Choice of kinetic term}

In the next sections, we will discuss two choices for $\cK$, one which enjoys Lorentz symmetry and another which does not. Other possibilities were discussed in \cite{revmu}. In both cases, assuming that the fractional term dominates in the UV, from \Eq{kei} one gets
\be
[\cK]_\textsc{ir}=2\,,\qquad [\cK]_\textsc{uv}=2\g\,,
\ee
which, combined with \Eq{scav}, yields the dimensionality of the scalar, which is given in the UV by
\be\label{phiuv}
[\phi]_\textsc{uv}=\frac{-[\rmd^Dx\,v]-[\cK]}{2}=\frac{D\a-2\g}{2}\,,
\ee
while in the IR one effectively recovers the energy dimension of an ordinary field:
\be
[\phi_{\rm eff}]_\textsc{ir}=\frac{D-2}{2}\,,\qquad \phi_{\rm eff}=\ell_*^{\frac{D\a-2\g}{2}-\frac{D-2}{2}}\phi\,.
\ee


\subsection{Power-counting renormalizability}\label{pca}

In order to get the superficial degree of divergence in a cut-off scheme, one can employ the usual power counting of ordinary QFT. This argument about renormalizability has been put forward from the onset of the multi-fractional proposal \cite{fra1,fra2} and repeated since then time and again \cite{revmu,frc2}. Let us revive it here, warning the reader that it will actually fail for the specific theories under inspection because it will grossly overestimate the actual degree of divergence of Feynman diagrams. 

For a non-trivial measure weight $v(x)\neq 1$ scaling as \Eq{scav}, there corresponds a multi-fractional measure in momentum space $\rmd^D k\,w(k)$ such that $[\rmd^D k\,w]=D\a$ in the UV. Consider a potential $V(\phi)=\la_N\phi^N$, where the coupling has energy dimension
\be
[\la_N]=D\a-N[\phi]\stackrel{\textrm{\tiny \Eq{phiuv}}}{=}\frac{2D\a-N(D\a-2\g)}{2}\,.
\ee
In a one-particle-irreducible graph with $L$ loops, $N_I$ internal propagators, $N_V$ vertices and $N_E$ external legs
, in the UV each momentum-space loop contributes $[\rmd^D k\,w]=D\a$ powers to the momentum integral defining the amplitude of the graph, while $[\tilde G]=-2\g$ for the propagator in momentum space (see below). Interaction vertices are dimensionless for the theory \Eq{actss} because the coupling $\la_N$ is a constant which does not contribute to the divergence of the integral. The divergent part of the graph in the UV scales as an energy cut-off $\Lambda_{\rm UV}$ to some power $\de$. Therefore, the superficial degree of divergence $\de$ in the UV is equal, by definition, to $[\textrm{loop}]L+[\textrm{propagator}]N_I=D\a L-2\g N_I$. Since $N_I\geq L$, then $\de\leq L(D\a-2\g)$ and the graph is convergent if $\g\geq D\a/2$. Also, using the topological relations $N N_V=N+2N_I$ and $L=N_I-N_V+1$, one ends up with
\be\label{sudedi}
\de = D\a L-2\g N_I =[\la_N](1-N_V)\,.
\ee
Since $N_V\geq 1$, to have $\de\leq 0$ the constant $\la_N$ must be positive semi-definite, which happens if 
\be
[\la_N]\geq 0\quad \Longleftrightarrow\quad \g\geq\frac{D\a}{2} \quad {\rm or}\quad N\leq \frac{2D\a}{D\a-2\g}\,.
\ee
In particular, for $\g=D\a/2$ the theory is power-counting renormalizable for any power $N$. We can summarize the results for multi-fractional theories:
\begin{itemize}
\item The theory $T_1$ has $\a\neq1$ and $\g=1$ (multi-scale measure and plain derivatives) and is power-counting renormalizable for $\a\leq 2/D$ but it is difficult to work out QFT in these spacetimes due to their lack of symmetries \cite{revmu}.
\item In the theory $T_v$ with weighted derivatives, $\a\neq 1$ and $\g=1$ (multi-scale measure and plain derivatives with weights). The power-counting argument does not show it but, in fact, the theory does not have improved renormalizability \cite{frc9}. 
\item The theory $T_q$ with $q$-derivatives corresponds to the case $\g=\a$ with $\a\neq 1$ (multi-scale measure and multi-scale derivatives). The superficial degree of divergence never becomes negative and it vanishes only in the delicate limit $\a\to 0$, which exists \cite{revmu} but goes beyond the scope of the present paper. Again, infinities in QFT are not tamed \cite{frc9}.
\item The class of theories $T_\g$ considered here have $\a=1$ and $\g\neq 1$ (plain measure and fractional operators), so that power-counting renormalizability is achieved when
\be\label{rg2}
\fl \boxd{\g\geq\frac{D}{2}} \quad {\rm or}\quad N\leq \frac{2D}{D-2\g}\qquad \stackrel{D=4}{\longrightarrow}\qquad \g\geq 2 \quad {\rm or}\quad N\leq \frac{4}{2-\g}\,.
\ee
For $\g=D/2$ ($=2$ in four dimensions) the theory is power-counting renormalizable (all couplings are positive) for any power $N$. As is well known, this theory with fourth-order derivatives is not unitary because it has a ghost; in the case of gravity, it corresponds to Stelle theory \cite{Ste77}. 
\end{itemize}


\section{Scalar theories with fractional derivatives}\label{sec3}

In this section, we outline the structure of the theories $T[\p^\g]$, $T[\p+\p^\g]$ and $T[\p^{\g(\ell)}]$.

To get a derivative operator with anomalous scaling, there are three options. One is to dress first- and second-order derivatives with measure factors; the theory $T_v$ with weighted derivatives $v^{-\b}\p_x(v^\b\,\cdot\,)$ and the theory $T_q$ with $q$-derivatives $\p_q:=v^{-1}\p_x$ are of this type, but they do not have improved renormalizability \cite{revmu}. Another possibility is to consider fractional derivatives \cite{MR,Pod99,SKM,KST}, which have numerous applications in engineering, percolation and transport theory, chaos theory, fractal geometry and complex systems \cite{frc1,fra6,frc4,MR,Pod99} and, in general, in any system with dissipation \cite{fra2,frc1} or a nowhere-differentiable geometry \cite{KoG}. The third option is to consider non-integer powers of the d'Alembertian; this will be done in section \ref{sec4}.

The theory $T[\p+\p^\g]$ is made of multi-fractional derivatives with explicit scaling, operators that we have to construct on demand for QFT applications \cite{mf0}. We will first study the theory $T[\p^\g]$ introducing fractional derivatives of a fixed order $\g$ in one dimension and then generalize them to $D$ dimensions and to operators with a scale dependence.


\subsection{Mixed fractional derivatives}\label{mixed}

We hereby introduce fractional derivatives of a single variable $x$. Partial fractional derivatives with spacetime indices $\mu$, $\nu$ will be considered starting from section \ref{tesec}.

The basic operators in fractional calculus are derivatives of non-integer order $\g$. There are many inequivalent ways to define these operators \cite{CaTe}. Here we select two specific choices, which are preferred among the others for various technical and physical reasons \cite{frc1}, mainly because we want the fractional derivative of a constant to be zero and we need to integrate on the whole axis. Namely, we choose the \emph{Liouville derivative}
\be\label{lio}
\fl {}_\infty\p^\g f(x) := \frac{1}{\Gamma(m-\g)}\int_{-\infty}^{+\infty}\rmd x'\, \frac{\Theta(x-x')}{(x-x')^{\g+1-m}}\p_{x'}^m f(x')\,,\qquad m-1\leq \g<m\,,
\ee
and the \emph{Weyl derivative}
\be\label{wey}
\fl {}_\infty\bar\p^\g f(x) := \frac{1}{\Gamma(m-\g)}\int_{-\infty}^{+\infty}\rmd x'\, \frac{\Theta(x'-x)}{(x'-x)^{\g+1-m}}\p_{x'}^m f(x')\,,\qquad m-1\leq \g<m\,,
\ee
where $\Theta$ is Heaviside's left-continuous step function:
\be\label{heaviside}
\Theta(x)=\cases{0, & $x \leq 0$ \\ 1, & $x > 0$}\,.
\ee
The Liouville derivative has memory of the past (integration from $-\infty$ to $x$), while the Weyl derivative encodes a pre-knowledge of the future (integration from $x$ to $+\infty$). Their properties, which can be found in \cite{MR,Pod99,SKM,KST,frc1}, are summarized in appendix \ref{appA}.

Fractional derivatives can be combined into the mixed operator \cite{frc1,Cre07,frc4}
\be\label{tD}
\cD^\g:=c\,{}_\infty\p^\g+\bar c\,{}_\infty\bp^\g\,.
\ee
Its properties can be derived from the elementary properties of fractional operators collected in appendix \ref{appA}. In the following, $-1=\rme^{\rmi\pi}$.
\begin{itemize}
\item[(a)] Limit to ordinary calculus (equations \Eq{limlio} and \Eq{limwey}):
\be\label{limD}
\lim_{\g\to n}\cD^\g =(c+\bar c \rme^{\rmi\pi n})\p^n\,,\qquad n\in\mathbb{N}.
\ee
\item[(b)] Linearity (equations \Eq{linlio} and \Eq{linwey}):
\be\label{linD}
\cD^\g[c_1 f(x)+c_2 g(x)]=c_1 (\cD^\g f)(x)+c_2(\cD^\g g)(x)\,.
\ee
\item[(c)] Composition rule (equations \Eq{comlio} and \Eq{comwey}): for all $\g,\b>0$,
\be\label{comD}
\fl \cD^\g\cD^\b = (c+\bar c)\cD^{\g+\b}+c\bar c({}_{\infty}\p^\g\,{}_{\infty}\bp^\b+{}_{\infty}\bp^\g\,{}_{\infty}\p^\b-{}_{\infty}\p^{\g+\b}-{}_{\infty}\bp^{\g+\b})\,.
\ee
\item[(d)] Kernel (equations \Eq{pllio} and \Eq{plwey}):
\be\label{plD}
\cD^\g x^\b = (c \rme^{\rmi\pi\g}+\bar c)\frac{\Gamma(\g-\b)}{\Gamma(-\b)} x^{\b-\g}.
\ee
Equation \Eq{plD} vanishes for $\b=0,1,2,\ldots,m-1$ and is ill-defined for $\b=\g$. The latter case can be treated separately. From the definitions of the operators ${}_\infty\p^\g$ and ${}_\infty\bp^\g$, for $m-1<\g<m$ the mixed fractional derivative of a polynomial $f_{m-1}$ of order $m-1$ is zero,
\be\label{1D}
\cD^\g f_{m-1}=0\,.
\ee
The mixed derivative of a constant is always zero and the kernel of a mixed derivative with $0<\g<1$ is trivial.
\item[(e)] Eigenfunctions (equations \Eq{explio} and \Eq{expwey}):
\be\label{expD}
\cD^\g \rme^{\rmi k x} =\left(c\,\rme^{\rmi\pi\frac{\g}{2}}+\bar c\,\rme^{-\rmi\pi\frac{\g}{2}}\right)k^{\g}\rme^{\rmi k x}\,,
\ee
where we used $\rmi=\rme^{\rmi\pi/2}$.
\item[(f)] Leibniz rule (equations \Eq{leru} and \Eq{lerub}):
\be\label{leruD}
\cD^\g(fg)=\sum_{j=0}^{+\infty}\frac{\Gamma(\g+1)}{\Gamma(\g-j+1)\Gamma(j+1)} (\p^j f)(\cD^{\g-j} g)\,.
\ee
A symmetric version of this rule exists for pure fractional derivatives \cite[equation (15.12)]{SKM} but its generalization to the mixed derivative $\cD^\g$ would require extra work not done here.
\item[(g)] Integration by parts (equation \Eq{ibp}). If $\bar c=\pm c^*$, then
\be\label{ibpD}
\int_{-\infty}^{+\infty}\rmd x\, f\,\cD^\g g = \pm\int_{-\infty}^{+\infty}\rmd x\, (\cD^{\g*} f)\,g\,.
\ee
\end{itemize}
 The choice
\be\label{ccbar}
c=\frac12\rme^{\rmi\theta}\,,\qquad \bar c=\pm\frac12\rme^{-\rmi\theta}\,,
\ee
where $\theta$ is a phase, clarifies the above formul\ae\ and eventually fixes $\theta$. In fact, calling $\cD^\g_\pm$ the mixed derivative with the $\pm$ choice in \Eq{ccbar}, equations \Eq{limD} and \Eq{expD} become
\ba
\lim_{\g\to n}\cD_+^\g &=& \cases{\cos\theta\,\p^n & $n$ even \\ \rmi\sin\theta\,\p^n & $n$ odd},\label{limD+}\\
\cD^\g_+ \rme^{\rmi k x} &=& \cos\left[\frac{\pi\g}{2}+{\rm sgn}(k)\,\theta\right]|k|^\g\,,\label{expD+}
\ea
and 
\ba
\lim_{\g\to n}\cD_-^\g &=&\cases{\rmi\sin\theta\,\p^n & $n$ even \\ \cos\theta\,\p^n & $n$ odd},\label{limD-}\\
\cD^\g_- \rme^{\rmi k x} &=& \rmi\,{\rm sgn}(k)\,\sin\left[\frac{\pi\g}{2}+{\rm sgn}(k)\,\theta\right]|k|^\g\,.\label{expD-}
\ea
Equations \Eq{limD+} and \Eq{limD-} are meaningful only for even $n$ or odd $n$, but not for all $n$ at the same time. In fact, in order to reproduce even-order or odd-order derivatives, one must rescale the operator $\cD^\g_\pm$ either by a factor $\cos\theta$ or by $\rmi\sin\theta$. Therefore, $\cD^\g_+$ and $\cD^\g_-$ are the generalization of, respectively, even and odd (or odd and even) integer derivatives. In particular, the mixed derivative with
\be\label{cc}
c=\bar c=\frac12\,,\qquad \theta=0\,,
\ee
is the generalization of even integer derivatives of order $n=0,2,4,\dots$, and \Eq{expD+} is
\be\label{expD3}
\cD^\g_+ \rme^{\rmi k x} =\cos\left(\frac{\pi\g}{2}\right)|k|^{\g}\rme^{\rmi k x}, \qquad\textrm{$n$ even}\,,
\ee
giving the correct limit $\g\to n$. When instead
\be\label{cc2}
c=-\bar c=\frac12\,,\qquad \theta=0\,,
\ee
one gets the generalization of odd integer derivatives of order $n=1,3,5,\dots$, and \Eq{expD-} becomes
\be\label{expD4}
\cD^\g_- \rme^{\rmi k x} = \rmi\,\sin\left(\frac{\pi\g}{2}\right)\,{\rm sgn}(k)|k|^\g, \qquad\textrm{$n$ odd}\,.
\ee
With these choices of phase and coefficients,
\be\label{tD+-}
\cD^\g_\pm=\frac12\left({}_\infty\p^\g\pm{}_\infty\bp^\g\right).
\ee

Note that $\cD^\g_\pm\cD^\g_\pm\neq \cD^{2\g}_\pm$ due to \Eq{comD}: 
\ba
\cD^\g_+\cD^\g_+ &=& \cD_+^{2\g}-\frac12(\cD_-^{2\g}-{}_{\infty}\p^\g\,{}_{\infty}\bp^\g)\neq \cD^{2\g}_+\,,\label{comD+}\\
\cD^\g_-\cD^\g_- &=& \frac12(\cD_+^{2\g}-{}_{\infty}\p^\g\,{}_{\infty}\bp^\g)\neq \cD^{2\g}_-\,,\label{comD-}
\ea
where we used \Eq{pllio} and \Eq{plwey} to find that ${}_{\infty}\p^\g\,{}_{\infty}\bp^\g={}_{\infty}\bp^\g\,{}_{\infty}\p^\g$.


\subsection{Theory \texorpdfstring{$T[\p^\g]$}{Tpg}}\label{tesec}

\subsubsection{Kinetic term and action.}

From the discussion of section \ref{mixed}, it becomes clear that the Liouville and Weyl derivative must coexist in the dynamics, since integrating by parts transforms one into the other. Therefore, we take a fractional kinetic term with mixed derivatives. The action \Eq{actss} with $v=1$ can be written in two ways:
\ba
\fl S_+&=&\int\rmd^Dx\left[\frac12\phi\,\eta^\mu\cD_{+\mu}^{2\g}\phi-V(\phi)\right],\label{actss2a}\\
\fl S_-&=&\int\rmd^Dx\left[\frac12\phi\,\eta^{\mu\nu}\cD_{-\mu}^\g\cD^\g_{-\nu}\phi-V(\phi)\right]=\int\rmd^Dx\left[-\frac12\eta^{\mu\nu}\cD_{-\mu}^\g\phi\,\cD^\g_{-\nu}\phi-V(\phi)\right]\!,\label{actss2b}
\ea
where $\mu=0,1,\dots, D-1$, $\eta^\mu=(-1,1,\dots,1)$, Einstein summation convention is used (in \Eq{actss2a}, $\cK=-\cD_0^{2\g}+\cD^{2\g}_1+\dots+\cD^{2\g}_{D-1}$) and we integrated by parts via \Eq{ibpD}. Equation \Eq{comD-} implies that \Eq{actss2a} and \Eq{actss2b} are inequivalent, since $\cD^\g_{-\mu}\cD^\g_{-\mu}\neq \cD^{2\g}_{+\mu}$ for any $\mu$. This can also be seen from \Eq{expD3} and \Eq{expD4}:
\ba
\fl \qquad\eta^\mu\cD_{+\mu}^{2\g} \rme^{\rmi k x} &=&\cos(\pi\g)|k|^{2\g}\rme^{\rmi k x}=:-b_+^\g|k|^{2\g}\rme^{\rmi k x}\,,\label{2plus}\\
\fl \eta^{\mu\nu}\cD_{-\mu}^\g\cD^\g_{-\nu} \rme^{\rmi k x} &=& -\sin^2\left(\frac{\pi\g}{2}\right)|k|^{2\g}= -\frac12[1-\cos(\pi\g)]|k|^{2\g}=:-b_-^\g|k|^{2\g}\rme^{\rmi k x}\,,\label{minmin}
\ea
where
\be\label{keke}
|k|^{2\g}:=-|k_0|^{2\g_0}+\sum_{i=1}^{D-1} |k_i|^{2\g}\,.
\ee

Guidance in the choice between \Eq{actss2a} and \Eq{actss2b} is given by the fact that our aim is to build a classical and quantum field theory encompassing both the matter and the gravity sector. Equation \Eq{actss2a}, however, does not allow for a generalization to a generic curved, non-diagonal metric $g^{\mu\nu}$, while \Eq{actss2b} does. The latter will be our pick. From now on, we will omit the $-$ subscript in the symbol $\cD$.

\subsubsection{Equation of motion.}

Variation of the action with respect to $\de\phi$ yields
\be\label{eomgf}
\eta^{\mu\nu}\cD_\mu^\g\cD_\nu^\g\phi-V'(\phi)=0\,,
\ee
where a prime denotes derivation with respect to $\phi$. Expanding the scalar field into Fourier modes, for the massive free case $V=m^{2\g}\phi^2/2$ we get the dispersion relation
\be\label{diregf}
(b_-^\g|k|^{2\g}+m^{2\g})\phi_k=0\,,
\ee
where $b_-^\g=[1-\cos(\pi\g)]/2$ and $m$ is a mass. Note that $0\leq b_-^\g\leq 1$ for all $\g$, while if we had chosen $b_+^\g$ we would have had windows in $\g$ where $b_+^\g<0$, with possible consequences on the unitarity of the theory.

\subsubsection{Dimensional flow.}\label{dimflo10}

In this theory, dimensional flow is trivial. Due to \Eq{v1}, the theory with multi-fractional derivatives lives on a spacetime with constant Hausdorff dimension both in position and in momentum space:
\be\label{dhgene}
\dh=\dh^k=D\,.
\ee
This means that rulers and clocks measure the same lengths and time intervals as in an ordinary setting. 

The spectral dimension is the dimension felt by a probe particle let diffusing in spacetime and it depends on the Hausdorff dimension of momentum space and on the kinetic term \cite{frc4,frc7,Calcagni:2019ngc}:
\be\label{dsgene} 
\ds=2\frac{\dh^k}{[\cK]}\,.
\ee
In the case of $T[\p^\g]$, the spectral dimension is constant and takes the same anomalous value at all scales:
\be\label{dsDg0}
\ds=\frac{D}{\g}\,.
\ee
Notice that $\ds$ is well-defined only when $\g>0$, since there is no physical meaning of negative dimensions.

In \cite{mf2}, we compare this and the other types of dimensional flow discussed in sections \ref{dimflo20} and \ref{dimflo30} with the dimensional flow found in quantum gravity.


\subsection{Theory \texorpdfstring{$T[\p+\p^\g]$}{Tppg}}

\subsubsection{Kinetic term and action.}

We define the multi-fractional derivative with explicit scaling as a linear combination of mixed fractional derivatives of different order:
\be\label{multider}
\cD_\mu:=\sum_\g u_\g\cD^\g_\mu\,.
\ee
The scaling is said to be explicit because the coefficients $u_\g$ are dimensionful and depend on one or more fundamental scales of spacetime geometry, such as $\ell_*$ introduced in \Eq{kei}. To recover standard QFT or gravity in some regime, we assume that $\g=1$ is included in the sum. 

Just like for the theory with fixed $\g$, we take the derivatives $\cD_-$ and a kinetic term $\phi\cK\phi$ in the action. Also, to make dimensional flow only one scale is sufficient \cite{revmu}, so that, overall, we choose as kinetic term $\cK$ the combination (we omit the $-$ subscript in $\cD$)
\be
\cK=\eta^{\mu\nu}\cD_\mu\cD_\nu\,.\label{profe}
\ee

Of all possible $D$-dimensional fractional derivatives, we choose the most symmetric one with $\g_\mu=\g$ for all $\mu$ because, otherwise, we would have to include extra scales in $\cD$. Another reason is that we want to treat all spacetime coordinates on the same ground, a sort of fractional covariance principle \cite{revmu}. However, later we will also consider a different realization where $\g_0=-\infty$ (no fractional derivative in the time direction):
\be\label{def2}
\cD_\mu = \ell_*^{1-\g}\p_\mu+\de_\mu^i\cD_\mu^\g\,,\qquad i=1,2,\dots,D-1\,,
\ee
corresponding to a fractional version of Ho\v{r}ava--Lifshitz field theories. The $k_0$ term is dropped in the definition of $|k|^{2\g}$ in \Eq{keke}.

\subsubsection{Equation of motion.}

Variation of the action with respect to $\de\phi$ yields
\be\label{eom}
\eta^{\mu\nu}\cD_\mu\cD_\nu\phi-V'(\phi)=0\,.
\ee
This equation is simpler than the most general one presented in \cite{frc2} for a non-trivial measure. Since plane waves are eigenfunctions both of the d'Alembertian $\B=\eta^\mu\p_\mu^2$ and of the mixed fractional derivatives, we can again expand in Fourier modes. In the free massive case,
\be\label{dire}
\left[\ell_*^{2(1-\g)}k^2+b_-^\g|k|^{2\g}+m^{2\g}\right]\phi_k=0\,,
\ee
where $b_-^\g$ is defined in \Eq{minmin} and $k^2:=k_\mu k^\mu=-k_0^2+\sum_{i=1}^{D-1} k_i^2$.

\subsubsection{Dimensional flow.}\label{dimflo20}

In the theory $T[\p+\p^\g]$ with the measure choice \Eq{v1}, the Hausdorff dimension is constant, equation \Eq{dsgene}, while the spectral dimension varies. In contrast, the Hausdorff dimension of the multi-fractional theory $T_v$ with weighted derivatives varies and the spectral dimension is constant, while both the spectral and Hausdorff dimensions vary in the multi-fractional theory $T_q$ with $q$-derivatives \cite{frc7}.

We can distinguish two types of geometry for the multi-fractional operator \Eq{profe}, plus another one for a generalization with two scales.
\begin{itemize}
\item When $\g>1$, the fractional operator in \Eq{profe} grows faster in momentum space than when $\g=1$ and it dominates the dynamics in the UV:
\be\label{dsuv}
\ds\stackrel{\textrm{\tiny UV}}{\simeq} \frac{D}{\g}\,,\qquad \ds\stackrel{\textrm{\tiny IR}}{\simeq} D\,,
\ee
so that there is a dimensional flow from an anomalous value at short scales to the topological dimension at large scales. Here `short' and `large' mean, respectively, below and above the microscopic scale $\ell_*$.
\item When $\g<1$, the spectral dimension is anomalous in the IR and standard in the UV:
\be\label{dsir}
\ds\stackrel{\textrm{\tiny UV}}{\simeq} D\,,\qquad \ds\stackrel{\textrm{\tiny IR}}{\simeq} \frac{D}{\g}\,.
\ee
For this theory, the quantum properties in the UV are the same as an ordinary QFT, so that there is no chance that renormalizability is improved. However, if $\ell_*$ is large enough the theory can work as a model with IR modifications which, in the case of gravity, can have interesting consequences at the cosmological level \cite{mf2}.
\item A third possibility is to extend the multi-fractional kinetic operator \Eq{profe} to three operators with three different exponents and two scales $\ell_1<\ell_2$:
\be\label{Kg30}
\cD_\mu = \p_\mu+\ell_1^{\g_1-1}\cD_\mu^{\g_1}+\ell_2^{\g_2-1}\cD_\mu^{\g_2}\,,\qquad \g_2<1<\g_1\,.
\ee
This dimensional flow has three regimes, a UV one where one can study the renormalizability of the theory, a mesoscopic one where standard classical and quantum field theory is recovered, and an IR or ultra-IR one which may have applications to cosmology:
\be\label{ds3reg}
\ds\stackrel{\textrm{\tiny UV}}{\simeq} \frac{D}{\g_1}<D\,,\qquad \ds\stackrel{\textrm{\tiny meso}}{\simeq} D\,,\qquad \ds\stackrel{\textrm{\tiny IR}}{\simeq} \frac{D}{\g_2}>D\,.
\ee
This is the most interesting case among the three but we will not study it in this paper, which is focussed on basic questions about the QFT in geometries of the first and second kind.
\end{itemize}


\subsection{Theory \texorpdfstring{$T[\p^{\g(\ell)}]$}{Tpgell}}

\subsubsection{Kinetic term and action.}

Another possibility is to define dimensional flow in terms of fractional derivatives of variable order, which corresponds to integrate over all scales \cite{frc2}. In the case where spacetime geometry has only one fundamental scale $\ell_*$,
\be\label{actvog}
S=\frac{1}{\ell_*}\int_0^{+\infty}\rmd\ell\,\tau(\ell)\int\rmd^Dx\left[\frac12\eta^{\mu\nu}\phi\cD_\mu^{\g(\ell)}\cD_\nu^{\g(\ell)}\phi-V(\phi)\right],
\ee
where $\ell$ is the probed scale and $\tau(\ell)$ is a dimensionless weight function.

\subsubsection{Equation of motion.}

Variation of the action with respect to $\de\phi$ yields
\be\label{eomgl}
\eta^{\mu\nu}\cD_\mu^{\g(\ell)}\cD_\nu^{\g(\ell)}\phi-V'(\phi)=0\,,
\ee
where we omit the integration in $\ell$ because the integrand should vanish at any given scale. Note that also the dimensionality of $\phi$ changes with the scale, according to our conventions for the weight $\tau(\ell)$.

The corresponding dispersion relation for the massive free case is
\be\label{diregl}
\left[b_-^{\g(\ell)}|k|^{2\g(\ell)}+m^{2\g(\ell)}\right]\phi_k=0\,,
\ee
where $b_-^{\g(\ell)}$ is defined in \Eq{minmin}, now with a scale dependence.

\subsubsection{Dimensional flow.}\label{dimflo30}

Just like for the theory $T[\p+\p^\g]$, we can distinguish three cases.
\begin{itemize}
\item With the profile \cite{frc2,fra6,frc4}
\be\label{Kellg}
\g(\ell)=\frac{1+(\ell_*/\ell)^2\g}{1+(\ell_*/\ell)^2}\,,
\ee
for any positive $\g$ one gets the dimensional flow \Eq{dsuv}. In the UV regime $\ell\ll\ell_*$, $\g(\ell)\to\g$, while in the IR regime $\ell\gg\ell_*$ one has $\g(\ell)\to 1$.
\item With the profile
\be\label{newprof}
\g(\ell)=\frac{\g+(\ell_*/\ell)^2}{1+(\ell_*/\ell)^2}\,,
\ee
for any positive $\g$ one recovers \Eq{dsir}.
\item With a profile $\g(\ell)$ \cite{fra6,frc4}
\be\label{prof2}
\g(\ell)=\frac{1+({\ell_1}/{\ell})^2\g_1+[{\ell_2}/({\ell-\ell_1})]^2\g_2}{1+({\ell_1}/{\ell})^2+[{\ell_2}/({\ell-\ell_1})]^2}\,,
\ee
with three plateaux $1$, $\g_1$ and $\g_2$ and two scales $\ell_1<\ell_2$, again we get the flow \Eq{ds3reg}, with the UV regime corresponding to $\ell\ll\ell_1$, the mesoscopic regime to $\ell\sim\ell_1$ and the IR regime to $\ell\gg\ell_2$. Here, however, we do not have to impose the condition $\g_2<1<\g_1$.
\end{itemize}


\subsection{Propagator and the quantum theory}

The propagator of the three theories can be read from the corresponding dispersion relation. Let us take the one for $T[\p+\p^\g]$, which has the most general structure. From \Eq{dire}, we get the bare Green function in momentum space
\be\label{prop}
\tilde G(k)=\frac{1}{\ell_*^{2(1-\g)}k^2+b_-^\g|k|^{2\g}+m^{2\g}}\,.
\ee
The same expression can be found rigorously from the Green's equation $\cK\,G(x)=-\de^D(x)$, as done in \cite{frc2} for non-mixed fractional derivatives.

The propagator \Eq{prop}, or its versions for $T[\p^\g]$ and $T[\p^{\g(\ell)}],$ is not analytic in $k^\mu$. The Osterwalder--Schrader conditions \cite{OsSc1,OsSc2}, necessary and sufficient for a Euclidean field theory to admit an analytic continuation to Lorentzian signature, include analyticity in the time component $k^0$ of the $D$-momentum. Thus, there is a tension between this requirement and reality of the kinetic operator spectrum that must be solved before quantizing the theory. We can do this in two ways.

\subsubsection{Efimov analytic continuation.}

The first solution to the analyticity problem is to define the theory with Efimov analytic continuation \cite{Efimov:1967dpd,Pius:2016jsl,Briscese:2018oyx}, used successfully in non-local quantum gravity \cite{Briscese:2018oyx}. As we mentioned in the introduction, in this prescription scattering amplitudes are calculated integrating along a sophisticated path $\cC$ in the complex plane of the energy $k^0$ circulating in the loop integrals. Through analytic continuation of the external momenta to purely imaginary energies $p^0=-\rmi p_D$, where $p_D\in\mathbb{R}$, the path $\cC$ is deformed so that $k^0\in\mathbb{I}$ is purely imaginary and one can define a purely real coordinate $k_D=\rmi k^0\in\mathbb{R}$ and integrate in $k_D$. After performing the integral, the external energy $p_D$ is analytically continued back to real values $p^0$ and one reconstructs the physical amplitude. In other words, one follows the same steps as in ordinary Lorentzian QFT: internal and external momenta are promoted to complex variables and a path $\cC$ is chosen to define the amplitude. The main difference between Efimov analytic continuation and traditional Wick rotation is that, in the first case, typical of non-local QFTs, the fields and amplitudes of the theory are defined at the very beginning with momentum integrals on the complex plane, along a path $\cC$ (or $\Gamma$, in the notation of \Eq{soluz0} below) that can be open or closed. While in the second case, typical of local QFTs and a special case of the former, fields and amplitudes are defined with Fourier transforms on the real line which are calculated by extending them to integrals on the complex plane on a path $\cC$. This path is actually a contour (i.e., it is closed) such that the arcs at infinity give a vanishing contribution and integration on the real axis $k^0\in\mathbb{R}$ can be rewritten as an integration on the imaginary axis $k^0\in\mathbb{I}$.

The main point of Efimov's procedure is that, since internal legs are described by Euclidean momenta, one can write the factor $|k|^{2\g}=\sum_{i=1}^D(k^2_i)^\g$ and perform all calculations in Euclidean space until the end, with no analyticity issue, while external momenta are in Lorentzian signature.

\subsubsection{Anisotropic derivatives.}

The analyticity problem can be bypassed also by considering ordinary time derivatives and fractional spatial derivatives, i.e., definition \Eq{def2}. In fact, the spatial momenta $k^i$ play the role of spectator parameters in the proof of the Osterwalder--Schrader conditions \cite{OsSc1,OsSc2} and analyticity of the propagator in $k^i$ is not required.

This is a sort of fractional extension of Ho\v{r}ava--Lifshitz gravity \cite{Hor3,Hor09,BPSi,HoMe,RACMP} where time derivatives are second-order and spatial derivatives are of higher but non-integer order \cite{Calcagni:2009qw}. Power-counting renormalizability has already been discussed and the scalar field theory is ghost free for the same reason as for the ordinary Ho\v{r}ava--Lifshitz scalar: the propagator has a simple mass pole with positive residue \cite{Hor09}. However, the absence of ghosts in the gravitational theory (present in some versions with ordinary derivatives) is much less obvious and may be worth a separate investigation.

\subsubsection{Other problems.}

The theories with fractional derivatives have two other problematic features that cannot be circumvented by Efimov analytic continuation or a choice of anisotropic derivatives.
\begin{itemize}
\item Although they are recovered in the IR, Lorentz and Poincaré invariance are broken in the UV and there is no other obvious symmetry replacing them. The absence of a rule can quickly lead the theory out of control in terms of naturalness of the Lagrangian, proliferation of operators, phenomenology, and so on. In particular, the Feynman rule for vertices is no longer a Dirac distribution of the sum of external momenta, which notably hinders the calculation of diagrams.
\item The Leibniz rule \Eq{leruD} may complicate otherwise elementary calculations already at the classical level. 
\end{itemize}
These issues add to the several technical points noted in \cite{frc1}, where a pessimistic view on the use of mixed derivatives was endorsed.


\section{Scalar theories with fractional d'Alembertian}\label{sec4}

We ended the previous section by listing three problems for the theory with fractional derivatives: non-analyticity of the propagator, lack of symmetries, and a complicated Leibniz rule. There is an easy way to solve the first and second problem at the same time. 


\subsection{Theory \texorpdfstring{$T[\B^\g]$}{TBg}}\label{sec4g}

\subsubsection{Kinetic term and action.}\label{kite1}

A derivative operator $\cK$ that scales anomalously, preserves Lorentz symmetry and can be readily generalized to a curved background is $\B^\g$, the non-integer power of the d'Alember\-tian. Allowing for a mass, we want to keep the pole structure as simple as possible and avoid complex poles. To that purpose, we incorporate the mass term inside the fractional operator
\be\label{Kgfix}
\cK= -(m^2-\B)^\g\,,
\ee
so that the action \Eq{actss} with $v=1$ reads
\be\label{actssfing}
\boxd{S=\int\rmd^Dx\left[-\frac12\phi(m^2-\B)^\g\phi-V(\phi)\right],}
\ee
where the potential includes only non-linear interactions. Of this theory, we will discuss the propagator, unitarity and renormalizability. Generalizing it to more fractional operators and fundamental scales is straightforward.

The fractional massive operator \Eq{Kgfix} can be written in a convenient position-space Schwinger representation \cite{Bal60}. Suppose $\g<n$, where $n\in\mathbb{N}$. 
 From the definition of the gamma function,
\ba
(m^2-\B)^\g&=&(m^2-\B)^{n}(m^2-\B)^{\g-n}\nonumber\\
&=&\frac{1}{\Gamma(n-\g)}\int_0^{+\infty}\rmd\tau\,\tau^{n-1-\g}\,(m^2-\B)^{n}\rme^{-\t(m^2-\B)}.\label{intpar}
\ea
One can use this formula to integrate by parts and derive the equation of motion $\de S/\de\phi=0$ from the action \Eq{actssfing}:
\be\label{scaomg}
(m^2-\B)^\g\phi+V'(\phi)=0\,.
\ee

\subsubsection{Dimensional flow.}\label{dimflo1}

The dimensional flow of this theory is trivial as in $T[\p^\g]$ (section \ref{dimflo10}). The Hausdorff dimension in position and momentum space is constant, \Eq{dhgene}, and so is the spectral dimension:
\be\label{dsDg}
\ds=\frac{D}{\g}\,.
\ee
This deviation from $D$ could violate experimental bounds on local spacetime geometry and particle-physics observables, unless $\g=1\pm\ve$ with $\ve\ll 1$. At present, we have no theoretical argument explaining why a fundamental theory would have such a fine tuning. We can still keep this as a possibility but in this paper we will mainly use the theory $T[\B^\g]$ as a spearhead to understand the other multi-fractional theories.

The spectral dimension is a meaningful geometric indicator only when $\g$ is positive,
\be\label{posibo}
\g>0\,,
\ee
a condition that will shrink the range on $\g$ allowed by unitarity (section \ref{unifrac}).

\subsubsection{Solutions of the free equation of motion.}\label{freeeq}

Let us now come to the problem of how to represent solutions of the free-field equation
\be\label{frefi}
\cK(\B)\,\phi(x)=0\,,
\ee
with a generic function $\cK$ of the d'Alembertian $\B$ in Lorentzian signature. We will follow \cite{BBOR1,BBOR2}, where the energy $k^0$ is analytically continued to the complex plane $(\Re\, k^0, \Im\, k^0)$. Decompose the scalar field into momentum modes,
\be
\hspace{-1cm}\phi(x)=\int_\Gamma\rmd^D k\,\rme^{-\rmi k\cdot x}\tilde\phi(k)\,,\qquad\qquad \int_\Gamma\rmd^D k:=\int_\Gamma\rmd k^0\int_{-\infty}^{+\infty}\rmd^{D-1} \bm{k}\,,\label{soluz0}
\ee
which looks similar to a Fourier transform but with the difference that integration in $k^0$ is not done along the real axis but on the path $\Gamma$, which runs from $-\infty$ to $+\infty$ along $\Im\, k^0>b$ and from $+\infty$ to $-\infty$ along $\Im\, k^0<-b$ for a given number $b\in\mathbb{N}$ (see figure \ref{fig1} below). $\tilde\phi(k)$ is analytic in the domain $\cC_b=\{k^0\,:\,|\Im k^0|>b\}$ and $\tilde\phi(k)/(k^0)^b$ is bounded continuous in the domain $\cC_b^==\{k^0\,:\,|\Im\, k^0|\geq b\}$.

Consider now a function $\cK(-k^2)$ analytic on the quotient domain $\cC_b/\cC_\b$ and such that $\cK(-k^2)/(k^0)^\b$ is bounded continuous in $\cC_b^=/\cC_\b^=$, for some $\b\in\mathbb{N}$. Then, applying $\cK(\B)$ to \Eq{soluz0} gives a well-defined expression:
\be\label{soluz1}
\cK(\B)\,\phi(x)=\int_\Gamma\rmd^D k\,\rme^{-\rmi k\cdot x}\cK(-k^2)\tilde\phi(k)\,,
\ee
where $\Gamma$ now runs from $-\infty$ to $+\infty$ for $\Im\, k^0>b+\b$ and from $+\infty$ to $-\infty$ for $\Im\, k^0<-b-\b$. Therefore, $\phi$ is a solution of the free-field equation of motion \Eq{frefi} when \Eq{soluz1} vanishes and this happens whenever the function $a(k):=-\cK(-k^2)\tilde\phi(k)$ is entire and analytic on the $(\Re\, k^0,\Im\, k^0)$ plane, so that by Jordan's lemma its integral on any closed path is zero, including on $\Gamma$:
\be\label{repra}
\int_\Gamma\rmd k^0\,\rme^{\rmi k^0 x^0}a(k^0,\bm{k})=0\,.
\ee
Note that this implies that $\cK^{-1}(-k^2)$ has the same analytic properties of $\tilde\phi(k)$. 

Define
\be\label{disfu}
\De_\pm(\cK):=\frac{1}{\cK[(k^0-\rmi\e)^2-|\bm{k}|^2]}\pm\frac{1}{\cK[(k^0+\rmi\e)^2-|\bm{k}|^2]}\,.
\ee
The path $\Gamma$ can be deformed around the singularities of $\cK^{-1}(-k^2)$ and can be split into two paths $\Gamma_-$ and $\Gamma_+$ sandwiched around the branch cuts on the real axis, if any, plus the loops $\Gamma_i$ circling around isolated singularities in the complex plane. In this way, the solution reads as the sum of the discontinuity functional $\De_-(\cK)$ \cite{BBOR1,BBOR2}, plus the contributions of the poles \cite{Belenchia:2014fda}:
\ba
\phi(x)&=&-\int_\Gamma\rmd^D k\,\rme^{-\rmi k\cdot x}\frac{a(k)}{\cK(-k^2)}\nonumber\\
       &=&\int_{-\infty}^{+\infty}\rmd^Dk\,\rme^{-\rmi k\cdot x}a(k)\,\Delta_-(\cK)-\sum_i\int_{\Gamma_i}\rmd^Dk\,\rme^{-\rmi k\cdot x}\frac{a(k)}{\cK(-k^2)},\label{soluz2}
\ea
where $\Gamma=\Gamma_+\cup\Gamma_-\bigcup_i\Gamma_i$. When all the singularities of the propagator $\cK^{-1}(-k^2)$ are on the real axis $\Im\, k^0=0$, the last contribution in \Eq{soluz2} vanishes and from the reality condition $a(-k)=a^*(k)$ we get
\be
\phi(x)=\int_{-\infty}^{+\infty}\rmd^Dk\,\left[a(k)\,\rme^{-\rmi k\cdot x}-a^*(k)\,\rme^{\rmi k\cdot x}\right]\Theta(k^0)\,\Delta_-(\cK).\label{soluz3}
\ee

In the specific case of the fractional kinetic term \Eq{Kgfix}, figure \ref{fig1} shows the contour $\Gamma$ and its deformation $\Gamma_+\cup\Gamma_-$ around the symmetric branch cut, where
\be\label{defom}
\om:=\sqrt{|\bm{k}|^2+m^2}\,.
\ee
Note that $\Delta_-(\cK)=0$ for $-\om\leq k^0\leq\om$.
\begin{figure}
\bc
\includegraphics[width=12cm]{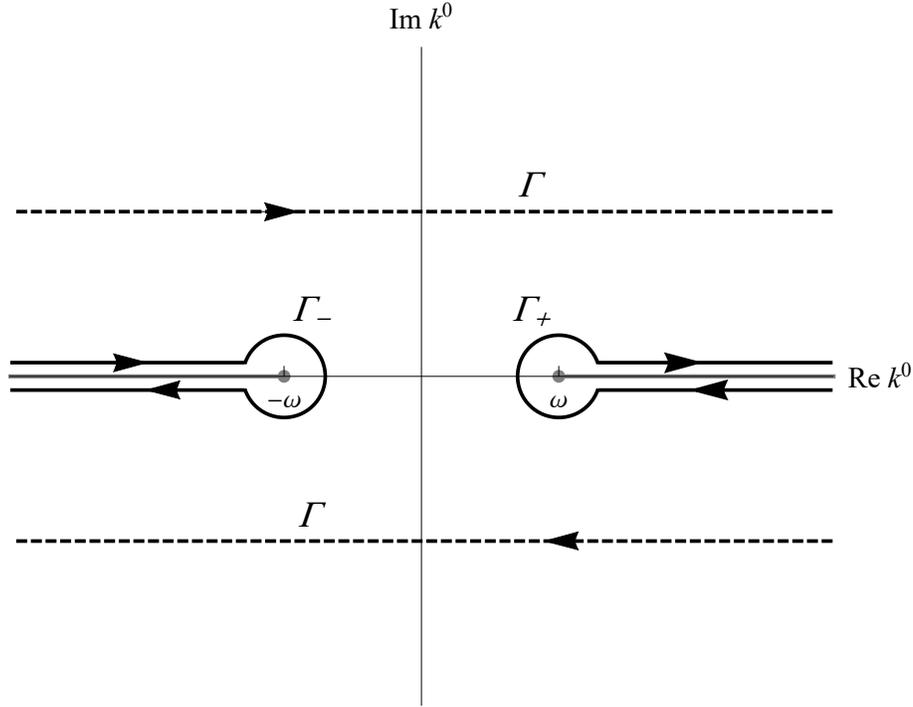}
\ec
\caption{\label{fig1} Contour $\Gamma$ (dashed lines) in the $(\Re\, k^0,\Im\, k^0)$ plane and its deformation $\Gamma_+\cup\Gamma_-$ (solid thick curves) around the branch cuts $k^0\leq -\om$ and $k^0\geq \om$ (gray thick lines).}
\end{figure} 

Defining the real-variable distributions \cite[section 3.2]{GeS}
\be
x_+^\la:=\Theta(x)\,x^\la\,,\qquad x_-^\la:=\Theta(-x)\,|x|^\la,
\ee
where $\Theta$ is defined in \Eq{heaviside} such that $\Theta(0)=0$, one can calculate the power of a complex variable when approaching the real axis from above and below \cite[section 3.6]{GeS}:
\be\label{geshi}
\lim_{\e\to 0}(x\pm\rmi\e)^{-\g}=x_+^{-\g}+\rme^{\mp\rmi\pi\g}x_-^{-\g}\,,
\ee
where $\g\neq 1,2,\dots$. Then,
\ba
\Delta_-(\cK)&=&[-(k^0+\rmi\e)^2+\om^2]^{-\g}-[-(k^0-\rmi\e)^2+\om^2]^{-\g}\nonumber\\
&=&[k^2+m^2-\rmi\e\, {\rm sgn}(k^0)]^{-\g}-[k^2+m^2+\rmi\e\, {\rm sgn}(k^0)]^{-\g}\nonumber\\
&=&{\rm sgn}(k^0)\left[(k^2+m^2-\rmi\e)^{-\g}-(k^2+m^2+\rmi\e)^{-\g}\right]\nonumber\\
&\stackrel{\textrm{\tiny \Eq{geshi}}}{=}& 2\rmi\,\sin(\pi\g)\,{\rm sgn}(k^0)\,(k^2+m^2)_-^{-\g}\nonumber\\
&=&2\rmi\,\sin(\pi\g)\,{\rm sgn}(k^0)\,\Theta(-k^2-m^2)\,|k^2+m^2|^{-\g}\\
&=&2\rmi\,\sin(\pi\g)\,{\rm sgn}(k^0)\,\Theta(k_0^2-\om^2)\,(k_0^2-\om^2)^{-\g}.\label{disfu2}
\ea
This weight function is spread on the domain $-k^2\geq m^2$, i.e., $k^0\in(-\infty,-\om]\cup[\om,+\infty)$, for a generic $\g\notin\mathbb{N}$, while for $\g\to 1$ one recovers the Lorentz-invariant free-wave solution with $\De_-(\cK)=2\pi\rmi\de(k^2+m^2)$, where we used the Sokhotski--Plemelj formula
\be\label{pleso}
\lim_{\e\to 0^+}\frac{1}{x-\rmi\e}={\rm PV}\left[\frac1x\right]+\rmi\pi\de(x)\,,
\ee
and PV denotes the principal value.

\subsubsection{Propagator.}\label{secpro}

The propagator of the free field is $-\rmi G(x)$, where the Green's function $G(x)$ is one of the solutions of the equation with source
\be\label{frefi2}
\cK(\B)\,G(x)=-\de^D(x)\,.
\ee
To solve it in momentum space, we need a function $f(k):=-\cK(-k^2)\tilde G(-k^2)$ that instead of \Eq{repra} yielded the delta distribution:
\be\label{repra2}
\int_\Gamma\rmd^D k\,\rme^{-\rmi k\cdot x}f(k)=\de^D(x)\,.
\ee
The spatial-momentum part is easy: since $(2\pi)^{D-1}\de^{D-1}(\bm{x})=\int_{-\infty}^{+\infty}\rmd^{D-1} \bm{k}\,\rme^{-\rmi \bm{k}\cdot \bm{x}}$, it follows that $f(k)=F(k^0)/(2\pi)^{D-1}$ for some function $F$. The latter is $F(k^0)={\rm sgn}(\Im\, k^0)/(4\pi)$ \cite{BOR}:
\ba
\frac12\int_\Gamma\frac{\rmd k^0}{2\pi}\,\rme^{\rmi k^0x^0}{\rm sgn}(\Im\, k^0) &=&\frac12\int_{-\infty}^{+\infty}\frac{\rmd k^0}{2\pi}\,\rme^{\rmi k^0x^0}-\frac12\int_{+\infty}^{-\infty}\frac{\rmd k^0}{2\pi}\,\rme^{\rmi k^0x^0}\nonumber\\
&=&\int_{-\infty}^{+\infty}\frac{\rmd k^0}{2\pi}\,\rme^{\rmi k^0x^0}=\de(x^0)\,.\nonumber
\ea
Therefore,
\ba
G(x) &=& -\frac12\int_\Gamma\frac{\rmd^D k}{(2\pi)^D}\,\rme^{-\rmi k\cdot x}\frac{{\rm sgn}(\Im\, k^0)}{\cK(-k^2)}\nonumber\\
&=& \frac12\int_{-\infty}^{+\infty}\frac{\rmd^Dk}{(2\pi)^D}\,\rme^{-\rmi k\cdot x}\De_+(\cK)\,,
\ea
where the two terms in $\De_+$ given in \Eq{disfu} are the causal and anti-causal Green's function, respectively.\footnote{In ordinary QFT, the causal and anti-causal Green's functions correspond, respectively, to the Feynman propagator and to the propagator encircling the poles in the opposite way compared to Feynman's. The latter is also called Dyson propagator \cite{GrRe} and it should not be confused with the one-loop-resummed propagator of section \ref{dypro}.} In particular, for the kinetic term \Eq{Kgfix}
\ba
\frac12\Delta_+(\cK)&=&\frac12\left\{[-(k^0+\rmi\e)^2+\om^2]^{-\g}+[-(k^0-\rmi\e)^2+\om^2]^{-\g}\right\}\nonumber\\
&=&\frac12\left\{[k^2+m^2-\rmi\e\, {\rm sgn}(k^0)]^{-\g}+[k^2+m^2+\rmi\e\, {\rm sgn}(k^0)]^{-\g}\right\}\nonumber\\
&=&\frac12\left[(k^2+m^2-\rmi\e)^{-\g}+(k^2+m^2+\rmi\e)^{-\g}\right]\nonumber\\
&\stackrel{\textrm{\tiny \Eq{geshi}}}{=}& (k^2+m^2)_+^{-\g}+\cos(\pi\g)(k^2+m^2)_-^{-\g}.\label{disfu3}
\ea

Other Green's functions can be obtained by adding a special solution to the solution of the homogeneous equation \cite{BBOR1} or, equivalently, by changing integration contour. For instance, one can take the causal or Feynman prescription \cite{doA92,BOR}:
\be\label{Gfey}
G_{\rm F}(x)=-\int_{\Gamma_{\rm F}}\frac{\rmd^D k}{(2\pi)^D}\,\rme^{-\rmi k\cdot x}\frac{1}{\cK(-k^2)}\,,
\ee
where $\Gamma_{\rm F}=\Theta(x^0)\,\Gamma_+\cup\Theta(-x^0)\,\Gamma_-$ and $\Gamma_\pm$ are shown in figure \ref{fig1}. The path $\Gamma_{\rm F}$ runs from below the branch cut $k^0\in(-\infty,\om)$ to above the cut $k^0\in(\om,+\infty)$. One can check that \Eq{Gfey} is a solution of  \Eq{frefi2} when applying $\cK(\B)$: the right-hand side yields $-(2\pi)^{-D}\int_{\Gamma_{\rm F}}\rmd^D k\,\rme^{-\rmi k\cdot x}=-(2\pi)^{-D}\int_{-\infty}^{+\infty}\rmd^D k\,\rme^{-\rmi k\cdot x}=-\de^D(x)$.

Two other possibilities are the advanced and retarded Green's functions $G_{{\rm R}, {\rm A}}(x)$ in position space for the fractional d'Alembertian. In the massless case ($m=0$) and up to a $(D,\g)$-dependent constant, one can check that \cite{doA92,BG}
\ba\label{Gretad}
G_{{\rm R}, {\rm A}}(t,r) &=& \int_{-\infty}^{+\infty}\frac{\rmd^D k}{(2\pi)^D}\,\rme^{-\rmi k\cdot x}\frac{1}{[-(k^0\pm\rmi\e)^2+\om^2]^\g}\nonumber\\
&\propto& \Theta(\mp t)\,\Theta(t^2-r^2)\,(t^2-r^2)^{\g-\frac{D}{2}},
\ea
where $t=x^0$ and $r^2=\sum_i x_i^2$. The intuitive origin of the power law is that this expression is the Fourier anti-transform of the inverse of the $\B^\g$ operator in momentum space, which is the power-law $\sim k^{2\g}$. In the limit $\g\to 1$ in $D=4$ dimensions, one recovers the standard result $G_{{\rm R}, {\rm A}}(t,r)\propto \Theta(\mp t)\, \de(t^2-r^2)=\de(r\pm t)/(2r)$ and the support of the advanced and retarded Green's functions is on the light cone. When $\g\neq 1$, however, the support spreads inside the light cone. This implies that virtual particles can travel slower, but no faster, than light.

Regardless of the specific choice of contour, the Green's function scales as
\be\label{propsi}
\tilde G(-k^2) =\frac{1}{(k^2+m^2)^\g}\,,
\ee
which has a branch cut $-k^2\geq m^2$ corresponding to time-like vectors with $(k^0)^2\geq \om^2$ (also light-like vectors if $m=0$). Thus, we cannot talk about a fundamental scalar particle because \Eq{propsi} has branch cuts instead of poles. In other words, while in ordinary QFT the bare propagator has poles and dressed propagators typically are non-local and contain branch cuts, 
 in theories with fractional operators the bare propagator itself can have branch cuts. The branch cut signals the presence of a continuum of modes with momenta $\leq -m$ and $\geq m$. In \cite{revmu}, these modes were described as `quasi-particles' in the lack of a better label. Perhaps, a characterization as a gas may also be a viable alternative.

Doing QFT with branch cuts in the bare propagator is possible: the structure of these objects is under control and has been studied in \cite{BOR,BBOR1,BBOR2,Belenchia:2014fda}. The appearance of branch cuts in quantum gravity and beyond the Einstein theory is not new, either, and people learned to live peacefully with them. Examples will be given in section \ref{actimm} and in \cite{mf2}.

\subsubsection{Unitarity: Källén--Lehmann representation.}\label{kalere}

The representation \Eq{intpar} suggests that the theory has ghosts due to the higher-derivative operator $(m^2-\B)^n$, while the exponential $\exp(\t\B)$ does not introduce any extra pole. One might then conclude that the quantum theory is unitary only if $n=1$, i.e., $\g<1$. However, in general integral parametrizations of non-local operators do not give direct, transparent information on the spectrum in non-local theories \cite{Calcagni:2018lyd,Calcagni:2018gke} and one should verify classical stability and quantum unitarity with other means. In this particular case, it will turn out that the unitarity bound $\g<1$ is almost correct.

In this sub-section, we will recall the basics of the Källén--Lehmann representation \cite{Kal52,Leh54,tV74b,Sre07,Zwi16}, an essential tool to verify unitarity of a QFT at all perturbative orders. At first, we will check only free-level unitarity, that is, the absence of ghosts in the free theory, which later we will extend to one-loop unitarity and, partially, to all loops.

\begin{figure}
\bc
\includegraphics[width=12cm]{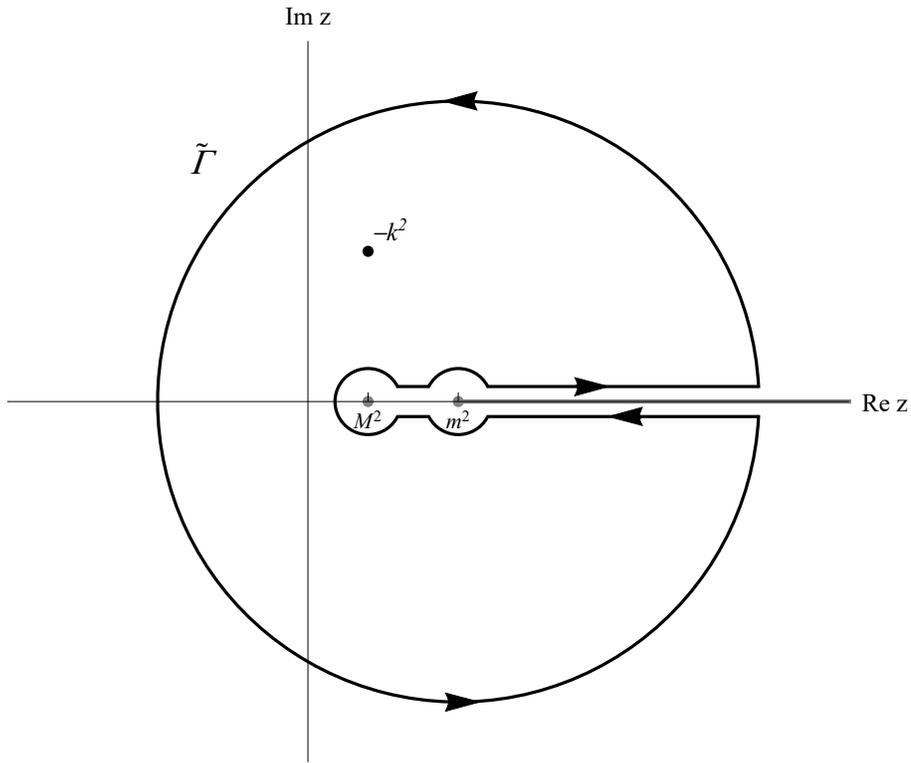}
\ec
\caption{\label{fig2} Contour $\tilde\Gamma$ (black thick curve) in the $(s=\Re\,z,\Im\,z)$ plane for a propagator with a simple pole at $z=M^2$ and a branch cut at $z\geq m^2$ (gray thick line).}
\end{figure}

Let $\tilde G(-k^2)$ be the Fourier transform of the exact (i.e., interacting) Green's function with Feynman prescription of a generic scalar field theory on Minkowski spacetime. Extending to the complex plane, assume that $\tilde G(z^*)=\tilde G^*(z)$ (this condition holds for our theory) and consider a closed contour $\tilde\Gamma$ encircling the point $z=-k^2$ and such that $\tilde G(z)$ is analytic inside and on $\Gamma$. Then, by Cauchy's integral formula the Green's function can be written as
\be\label{opt}
\tilde G(-k^2)=\frac{1}{2\pi\rmi}\oint_{\tilde\Gamma}\rmd z\,\frac{\tilde G(z)}{z+k^2}\,.
\ee
Suppose that $\tilde G$ is singular at several places on the real axis. For instance, if $\tilde G(z)$ had a simple pole at $z=M^2$ and a branch point at $z=m^2>M^2$, with a branch cut on the positive $z\geq m^2$ half line, the contour would be the one in figure \ref{fig2}.\footnote{Since $\tilde G(z)$ is analytic on the contour path, the integrand in \Eq{opt} has a simple pole at $z=-k^2$ and the residue theorem gives $+2\pi\rmi \tilde G(-k^2)$, with the contour $\tilde\Gamma$ in the counter-clockwise direction.}

\begin{figure}
\bc
\includegraphics[width=12cm]{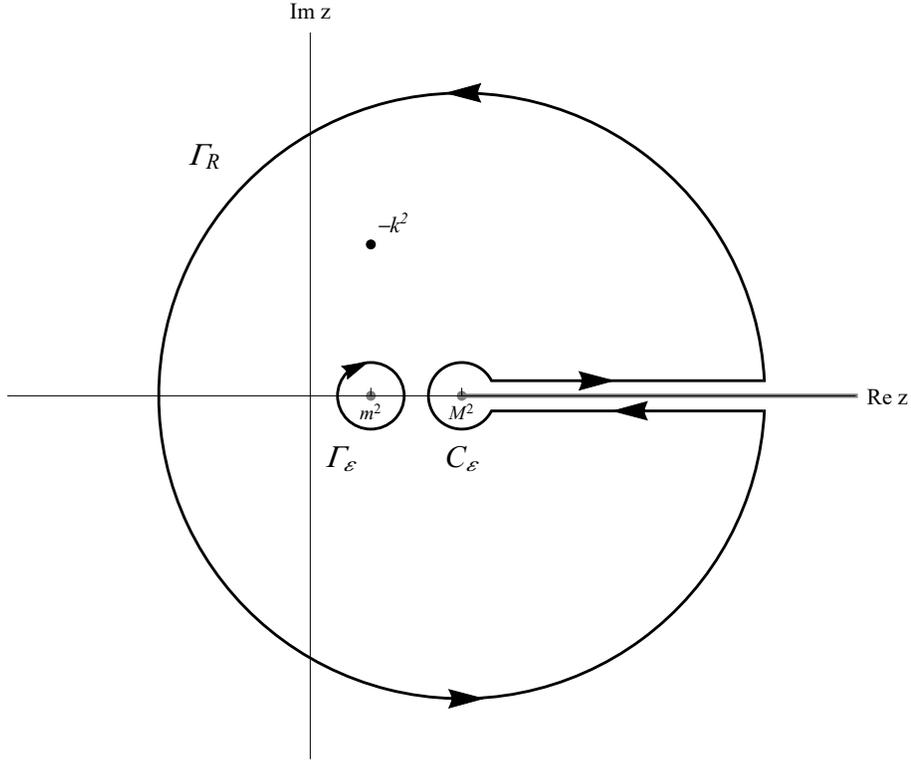}
\ec
\caption{\label{fig3} Deformation of the contour $\tilde\Gamma$ of figure \ref{fig2} into two disconnected contours (black thick curves), for a propagator with a simple pole at $z=M^2$ and a branch cut at $z\geq m^2$ (gray thick line).}
\end{figure}

We can deform the contour continuously as in figure \ref{fig3} and split it into four parts: a loop $\Gamma_\ve$ of radius $\ve$ encircling the simple pole $z=M^2$, a mini-contour $C_\ve$ of radius $\ve$ around the branch point $z=m^2$, an empty contour in the region $M^2<z<m^2$ not shown in the figure, the paths going back and forth along the branch cut and a counter-clockwise circle $\Gamma_R$ of radius $R$.

The contribution of the empty contour to \Eq{KLrepr} is zero by the Cauchy--Goursat theorem\footnote{$\tilde G(s)$ is monodromic and analytic for $M^2<s<m^2$ and there is no discontinuity when crossing the real axis $\tilde G(s+\rmi\e)=\tilde G(s-\rmi\e)$, hence $\Im\,\tilde G(s+\rmi\e)=0$ in this region.} and so is the contribution of $\Gamma_R$ when $R\to\infty$ if $\tilde G(z)$ falls off at $z\to\infty$. Parametrizing $z$ as $z=m^2+\ve\exp(\rmi\theta)$ with $-2\pi<\theta<0$, the contribution of the contour around the branch point is
\be\label{ceps}
\frac{1}{2\pi\rmi}\int_{C_\ve}\rmd z\,\frac{\tilde G(z)}{z+k^2} = \frac{\ve}{2\pi(k^2+m^2)}\int_{-2\pi}^0\rmd\theta\,\rme^{\rmi\theta}\tilde G(m^2+\ve\,\rme^{\rmi\theta})+\dots\,,
\ee
where the ellipsis stands for higher-order terms in $\ve$. This integral can vanish, diverge or be finite depending on $\tilde G$. Assuming that it vanishes (which must be checked explicitly for any given $\tilde G$), the only contributions left are those of the pole at $z=M^2$ and of the paths along the branch cut. The latter is
\ba
\frac{1}{2\pi\rmi}\int_{\rm cut}\rmd z\,\frac{\tilde G(z)}{z+k^2} &=&\lim_{\e\to0^+}\frac{1}{2\pi\rmi}\int_{m^2}^{+\infty}\rmd s\,\frac{\tilde G(s+\rmi\e)-\tilde G(s-\rmi\e)}{k^2+s-\rmi\e}\nonumber\\
&=&\frac{1}{\pi}\lim_{\e\to0^+}\int_{m^2}^{+\infty}\rmd s\,\frac{\Im[\tilde G(s+\rmi\e)]}{k^2+s-\rmi\e}\nonumber\\
&=& \int_{m^2}^{+\infty}\rmd s\,\frac{\rho(s)}{k^2+s-\rmi\e}\,,\label{KLrepr}
\ea
where $s=\Re\,z$ and\footnote{With an abuse of notation, we leave the contour prescription $-\rmi\e$ in \Eq{KLrepr} out of the limit $\e\to 0^+$ taken in \Eq{rhoG}.}
\be\label{rhoG}
\rho(s):=\frac{1}{\pi}\,\lim_{\e\to 0^+}\Im[\tilde G(s+\rmi\e)]\,.
\ee
The loop around the pole admits a similar expression because one can divide $\Gamma_\ve$ into two lines above and below the real axis plus two infinitesimal half arcs that give a zero contribution:
\be\label{polerepr}
\frac{1}{2\pi\rmi}\int_{\Gamma_\ve}\rmd z\,\frac{\tilde G(z)}{z+k^2}= \int_{M^2-\ve}^{M^2+\ve}\rmd s\,\frac{\rho(s)}{k^2+s-\rmi\e}\,.
\ee
By construction, the support of $\de(s+k^2)$ is to the right of the lower integration limit $s_{\rm min}=M^2-\ve$. 

Combining \Eq{KLrepr} and \Eq{polerepr}, the Källén--Lehmann representation is \cite{Kal52,Leh54,tV74b}
\be\label{KLrepr2}
\tilde G(-k^2) =\int_{M^2-\ve}^{M^2+\ve}\rmd s\,\frac{\rho(s)}{k^2+s-\rmi\e}+\int_{m^2}^{+\infty}\rmd s\,\frac{\rho(s)}{k^2+s-\rmi\e}\,.
\ee
From the representation \Eq{KLrepr2} for the exact propagator, one can determine the spectral function via the Sokhotski--Plemelj formula \Eq{pleso}. 

For the theory to be unitary, the spectral function must be positive semi-definite for all $s$ in the integration domain:
\be\label{rhos}
\rho(s)\geq 0\,.
\ee
As we will see in appendix \ref{appB} for the scalar theory $T[\B^\g]$, this condition is equivalent to impose reflection positivity, one of the Osterwalder--Schrader conditions \cite{OsSc1,OsSc2} required for a Euclidean field theory to admit an analytic continuation to Minkowski spacetime. If reflection positivity is violated, there is no positive semi-definite scalar product in the space of functionals of the field $\phi$ and there are no unitary representations of the Poincaré group. This would signal the presence of negative-norm states (ghosts).

\subsubsection{Unitarity: ordinary theory.}\label{orsc}

For the ordinary interacting scalar field theory \Eq{ordi}, where $\cK=\B$, the full propagator has the singularities shown in figure \ref{fig2}, a simple pole at $z=M^2$ and a branch cut at $z\geq m^2=4M^2$, where $m^2$ is the lowest mass in the multi-particle spectrum. 

In the presence of interactions, the first contribution is the renormalized free part, while the second encodes multi-particle states.\footnote{In \cite{Sre07}, the spectral function is defined as the interacting part of our $\rho$, while the latter coincides with the free$+$interacting $\rho$ of \cite{Zwi16}.} Here we will be interested only in the free part without interactions. The free propagator in Minkowski momentum space is
\be\label{freeG}
\tilde G(-k^2) = \frac{1}{k^2+M^2}\qquad\Rightarrow\qquad \tilde G(s+\rmi\e) = -\frac{1}{s-M^2+\rmi\e}\,,
\ee
where $\e>0$. The determination of the free spectral function is an almost tautological exercise because we can already read off the sign of the residue from the propagator, but we will do it anyway because we are interested in the parallelism with the fractional case. In fact, in the ordinary case one can check the absence of ghosts from the rule-of-the-thumb `the sign of the residue (of the propagator) must be positive,' but in the fractional case there is no residue to begin with.

A quick way to find $\rho(s)$ is to multiply and divide \Eq{freeG} by $k^2+M^2+\rmi\e$, so that from \Eq{rhoG}
\ba
\rho(s)&=&\lim_{\e\to0^+}\frac{1}{\pi}\,\Im\!\left[-\frac{s-M^2-\rmi\e}{(M^2-s)^2+\e^2}\right]=\lim_{\e\to 0^+}\frac{1}{\pi}\frac{\e}{(s-M^2)^2+\e^2}\nonumber\\
&=&\de(s-M^2)\,,\label{poi}
\ea
where we used the representation of the Dirac delta distribution as the limit of the Poisson kernel. When $s\neq M^2$, the limit when $\e\to 0^+$ is zero, while when $s=M^2$ it diverges as $1/(\pi\e)$. This is the behaviour of the Dirac distribution.

The same result can be reached from the Schwinger representation of \Eq{freeG} in Lorentzian signature:
\be\label{schw}
\fl \tilde G(-k^2) = \rmi\int_0^{+\infty}\rmd\tau\,\rme^{-\rmi\tau(k^2+M^2-\rmi\e)}\quad\Rightarrow\quad \tilde G(s+\rmi\e) = \rmi\int_0^{+\infty}\rmd\tau\,\rme^{-\rmi\tau(M^2-s-\rmi\e)}\,.
\ee
From \Eq{schw},
\ba
\rho(s) &\stackrel{\textrm{\tiny \Eq{rhoG}}}{=}&\frac{1}{\pi}\,\lim_{\e\to 0^+}\int_0^{+\infty}\rmd\tau\,\rme^{-\tau\e}\cos[\tau(s-M^2)]\nonumber\\
&=& \lim_{\e\to 0^+}\frac{1}{\pi}\frac{\e}{(s-M^2)^2+\e^2}=\de(s-M^2)\,.\nonumber
\ea
Thus, the spectral distribution is positive semi-definite and singular at $s=M^2$, corresponding to the $k_0^2=\om^2$ mass pole. Integrating between $M^2-\ve$ and $M^2+\ve$, one recovers the Green's function \Eq{freeG}:
\be\nonumber
\int_{M^2-\ve}^{M^2+\ve}\rmd s\,\frac{\de(s-M^2)}{k^2+s-\rmi\e}=\frac{1}{k^2+M^2-\rmi\e}\,.
\ee
In this case, there is a neat correspondence between positivity of the pole residue and positivity of the spectral function (absence of ghosts).

\subsubsection{Unitarity: fractional theory.}\label{unifrac}

\begin{figure}
\bc
\includegraphics[width=12cm]{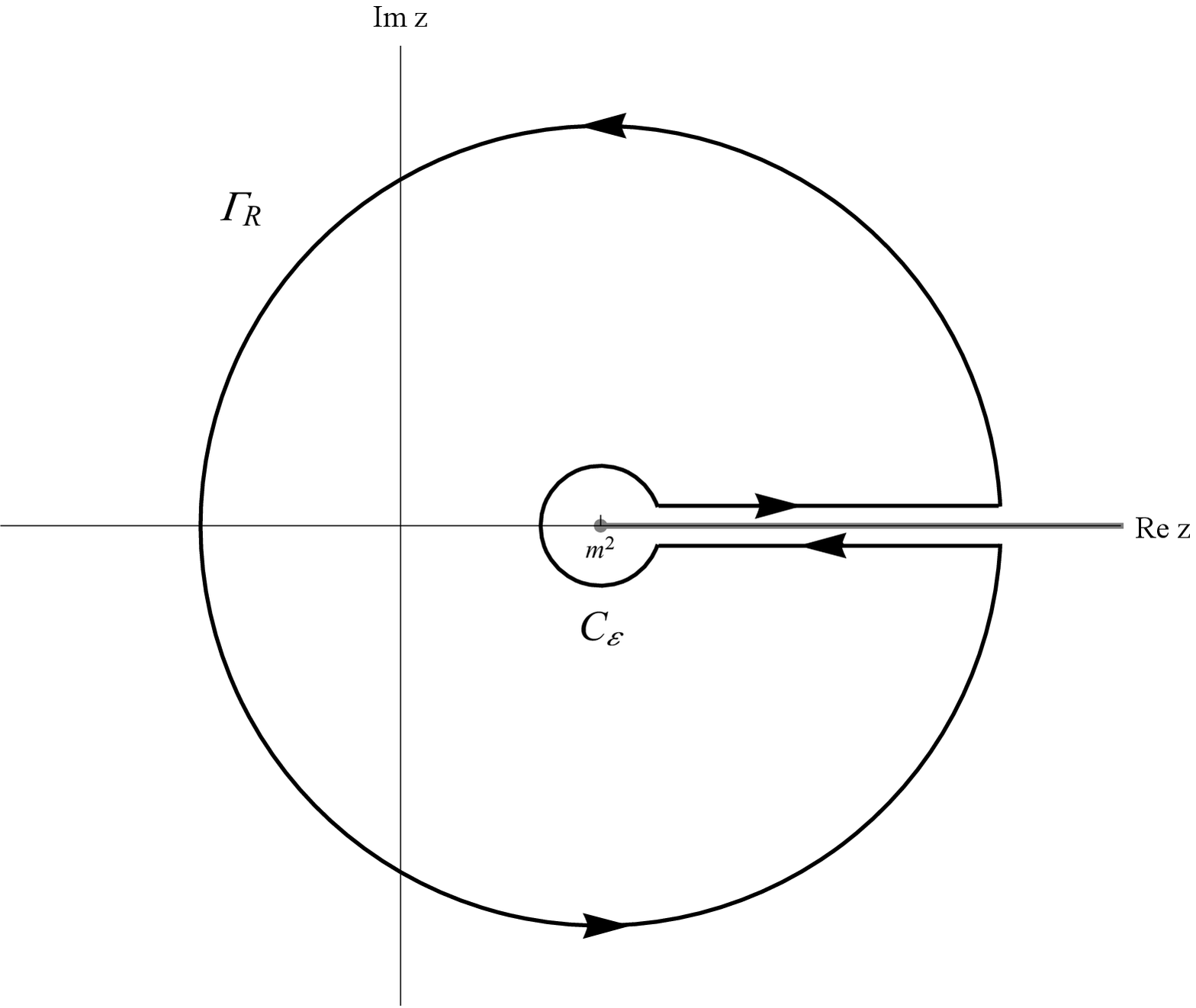}
\ec
\caption{\label{fig4} Contour $\tilde\Gamma$ (black thick curve) in the $(s=\Re\, z,\Im\, z)$ plane for a propagator with a branch cut at $z\geq m^2$ (gray thick line).}
\end{figure}

Let us repeat the procedure of section \ref{orsc} for the fractional causal Green's function
\be\label{Gfrac}
\tilde G(-k^2) = \frac{1}{(k^2+m^2)^\g}\,,\qquad\Rightarrow\qquad \tilde G(s+\rmi\e)=\frac{1}{(m^2-s-\rmi\e)^\g}\,,
\ee
where the prescription $-\rmi\e$ is such that \Eq{Gfrac} yields the Feynman propagator in the limit $\g\to 1$. This function has a branch cut at $s\geq m^2$ and no isolated poles (figure \ref{fig4}). First of all, we check for which $\g$ the piece of contour around $z=m^2$ gives a vanishing contribution in the limit $\ve\to 0^+$. Plugging \Eq{Gfrac} into \Eq{ceps},
\ba
\frac{1}{2\pi\rmi}\int_{C_\ve}\rmd z\,\frac{\tilde G(z)}{z+k^2} &=& \frac{\ve^{1-\g}\rme^{-\rmi\pi\g}}{2\pi(k^2+m^2)}\int_{-2\pi}^0\rmd\theta\,\rme^{\rmi\theta(1-\g)}+O(\ve^{2-\g})\nonumber\\
&=&-\frac{\ve^{1-\g}}{(1-\g)(k^2+m^2)}\frac{\sin(\pi\g)}{\pi} +O(\ve^{2-\g})\,,
\ea
which implies
\be\label{rg0}
\g<1\,.
\ee
Then, the Källén--Lehmann representation of the Green's function is
\be\label{KLrepr3}
\tilde G(-k^2) = \int_{m^2}^{+\infty}\rmd s\,\frac{\rho(s)}{k^2+s-\rmi\e}\,.
\ee

The first method to compute the spectral function, analogous to \Eq{poi}, is to write
\ba
\tilde G(s+\rmi\e) &=& (m^2-s-\rmi\e)^{-\g}=\exp\left[-\g\,{\rm Ln}(m^2-s-\rmi\e)\right]\nonumber\\
		 &=& \exp\left[-\g \ln\sqrt{(s-m^2)^2+\e^2}-\rmi\g\,{\rm Arg}(m^2-s-\rmi\e)\right],
\ea
where Ln and Arg are the principal value of the complex logarithm and of the argument (the phase of a complex number). Then,
\be\nonumber
\rho(s)=-\lim_{\e\to 0^+}\frac{1}{\pi}\frac{1}{[(s-m^2)^2+\e^2]^{\g/2}}\sin\left[\g\,{\rm Arg}(m^2-s-\rmi\e)\right].
\ee
In the presence of a branch cut on the real axis, the principal value Arg is evaluated quadrant by quadrant separately, depending on the sign of the real part $m^2-s$ and of the imaginary part $\e$. In our case, $m^2-s<0$ and $-\e\leq 0$, which implies that
\be\nonumber
{\rm Arg}(m^2-s-\rmi\e)={\rm arctan}\left(\frac{\e}{m^2-s}\right)-\pi\,,
\ee
so that
\ba
\hspace{-1.5cm} \rho(s)&=&-\lim_{\e\to 0^+}\frac{1}{\pi}\frac{1}{[(s-m^2)^2+\e^2]^{\g/2}}\sin\left[\g\,{\rm arctan}\left(\frac{\e}{m^2-s}\right)-\pi\g\right]\label{rhofin0}\\
\fl &=&\frac{\sin(\pi\g)}{\pi}\frac{1}{(s-m^2)^\g}\,.\label{rhofinal}
\ea
 The same result comes from the expression 
\be\label{Schwg}
\tilde G(s+\rmi\e)=\frac{\rme^{\rmi\frac{\pi\g}{2}}}{\Gamma(\g)}\int_0^{+\infty}\rmd\tau\,\tau^{\g-1}\rme^{-\rmi\tau(m^2-s-\rmi\e)}\,,
\ee
where we generalized the Schwinger representation \Eq{schw} by using the definition of the gamma function $\Gamma(\g)=\int_0^{+\infty}\rmd x\,x^{\g-1}\rme^{-x}$, valid for $\g>0$ but that can be analytically continued to all $\g\neq 0,-1,-2,\dots$.
Taking the imaginary part and integrating in $\tau$,
\ba
\fl \rho(s) &\stackrel{\textrm{\tiny \Eq{rhoG}}}{=}&\frac{1}{\pi\Gamma(\g)}\,\lim_{\e\to 0^+}\int_0^{+\infty}\rmd\tau\,\tau^{\g-1}\rme^{-\tau\e}\nonumber\\
\fl &&\qquad\qquad\qquad\times\left\{\sin\left(\frac{\pi\g}{2}\right)\,\cos[\tau(m^2-s)]-\cos\left(\frac{\pi\g}{2}\right)\,\sin[\tau(m^2-s)]\right\}\nonumber\\
\fl &=&\frac{1}{\pi\Gamma(\g)}\,\lim_{\e\to 0^+}\int_0^{+\infty}\rmd\tau\,\tau^{\g-1}\rme^{-\tau\e}\sin\left[\frac{\pi\g}{2}+\tau(s-m^2)\right]\nonumber\\
\fl &=& \lim_{\e\to 0^+}\frac{1}{\pi[(s-m^2)^2+\e^2]^{\g/2}}\sin\left(\frac{\pi\g}{2}-\g\,{\rm arctan}\frac{m^2-s}{\e}\right)\nonumber\\
\fl &=&\frac{\sin(\pi\g)}{\pi}\frac{1}{(s-m^2)^\g}\,,\nonumber
\ea
where we used the fact that $\lim_{\e\to0^+}{\rm arctan}[(m^2-s)/\e]=-\pi/2$, since $m^2-s<0$. The last line is \Eq{rhofinal}; integrating it in \Eq{KLrepr3}, one recovers \Eq{Gfrac}. Note that, just like the limit $s\to m^2$, the limit $\g\to 1$ does not commute with the limit $\e\to 0^+$ and one can recover the delta distribution \Eq{poi} only from \Eq{rhofin0}, not from \Eq{rhofinal} (a similar non-commutation rule holds, for instance, in causal sets \cite{Belenchia:2014fda}).

The spectral function \Eq{rhofinal} is real ($s>m^2$) and is positive definite when $\sin(\pi\g)> 0$. This condition and inequality \Eq{rg0} fix the allowed range for $\g>0$ to $0<\g<1$, while for $\g<0$ one has $-2<\g<-1$, $-4<\g<-3$, and so on:
\be\label{rg1}
\boxd{-2n<\g<1-2n\leq 1\,,\qquad n\in\mathbb{N}\,.}
\ee
We excluded equalities in order to get a non-trivial function $\rho\neq 0$ but later they will be removed anyway imposing one-loop finiteness. 
 In appendix \ref{appB}, we recover the unitarity constraint \Eq{rg1} with a different method based on reflection positivity.

The lowest interval ($n=0$) 
\be\label{rg1min}
0<\g<1
\ee
is the only one for which the spectral dimension of spacetime is positive definite, according to \Eq{dsDg}. Note that the ranges \Eq{rg1} correspond to a theory with modifications dominating in the IR.

\subsubsection{One-loop renormalization: vertices and vacuum diagram.}\label{1lreno1}

In section \ref{pca}, we have found that the theory is power-counting renormalizable for $\g\geq D/2$, corresponding to $\g\geq 2$ in four dimensions. The power-counting argument holds in a cut-off regularization scheme and one should wonder whether loop integrals contain other divergences than those seen in this scheme. Here we compute two one-loop diagrams for the $\phi^3$ theory, the vacuum diagram and the self-energy, to show that they are finite.

Each internal leg contributes a factor $-\rmi \tilde G(-k^2)$, each loop corresponds to an integration in $\rmd^Dk$ and each bare vertex is identical to the ordinary lowest-order $N$-particle amplitude:
\ba
\parbox{2.5cm}{\includegraphics[width=2.3cm]{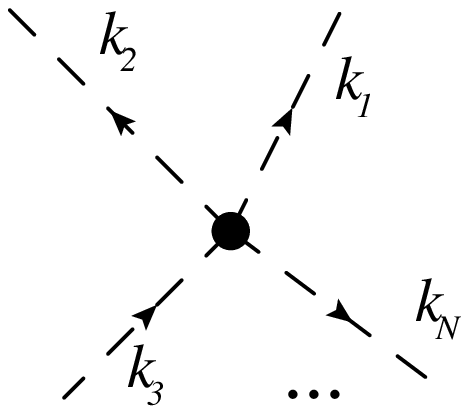}}&=&\cV(k_1,\dots,k_N)=\rmi\la_N\int\rmd^D x\,\rme^{\rmi k_{\rm tot}\cdot x}\nonumber\\
&=&\rmi\la_N(2\pi)^D\de^D(k_{\rm tot})\,,\label{vert}
\ea
where $k^\mu_{\rm tot}:=\sum_{n=1}^N k_n^\mu$. In the following, $N=3$ and we use the standard notation for scalar-field diagrams: dashed lines and, for vertices, thick dots as in \Eq{vert}. Some of the integrals involving \Eq{Gfrac} can be found as limits of the formul\ae\ in \cite[appendix A]{tV72}.

Vacuum diagrams have no external legs. At one loop, the only contribution is the one-point function, a propagator closing on itself:
\be
\parbox{1.6cm}{\includegraphics[width=1.5cm]{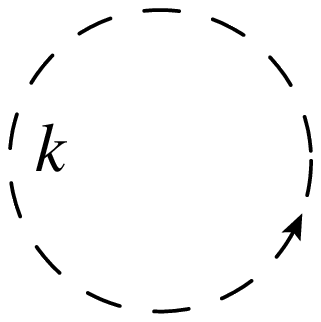}}=\rmi\cA=\int\frac{\rmd^D k}{(2\pi)^D}\,\frac{-\rmi}{(k^2+m^2-\rmi\e)^\g}\,.\label{vacd}
\ee
To compute it, we analytically continue to Euclidean momentum space and use the Euclidean version of the Schwinger representation \Eq{Schwg}, with $\tau=-\rmi\s$:
\ba
\rmi\cA&=&\frac{1}{\Gamma(\g)}\int\frac{\rmd^D k}{(2\pi)^D}\int_0^{+\infty}\rmd \s\,\s^{\g-1}\rme^{-\s(k^2+m^2)}\nonumber\\
&=&\frac{\Om_D}{(2\pi)^D\Gamma(\g)}\int_0^{+\infty}\rmd \s\,\s^{\g-1}\int_0^{+\infty}\rmd k\,k^{D-1}\,\rme^{-\s(k^2+m^2)}\nonumber\\
&=&\frac{1}{2^D\pi^{\frac{D}{2}}\Gamma(\g)}\int_0^{+\infty}\rmd \s\,\s^{\g-1-\frac{D}{2}}\,\rme^{-m^2\s}\nonumber\\
&=&\frac{\Gamma\left(\g-\frac{D}{2}\right)}{2^D\pi^{\frac{D}{2}}\Gamma(\g)}(m^2)^{\frac{D}{2}-\g}\,,\label{vacd2}
\ea
where $\Om_D=2\pi^{D/2}/\Gamma(D/2)$ is the surface of the unit $D$-ball coming from integration of the solid angle in momentum polar coordinates, $\rmd^Dk=\rmd\Om_D\rmd k\,k^{D-1}$. Note that integration in $\s$ and $k$ commutes. 

When $\g=1$, equation \Eq{vacd2} agrees with the ordinary scalar-field one-point function found in dimensional regularization \cite{tV79}. In that case, the result diverges like $\Gamma(-1)$ in $D=4$ and a subtraction scheme must be enforced. In our case, the result is finite if
\be
\boxd{\g-\frac{D}{2}\neq -n=0,-1,-2,\dots\,,}
\ee
which corresponds to $\g\neq 2,1,0,-1,-2,\dots$ in four dimensions. These values are already excluded by the lower bound \Eq{rg2} for power-counting renormalizability.

\subsubsection{One-loop renormalization: self-energy diagram.}\label{1lreno2}

In the following, we rescale the coupling constant as $\lambda_N\to\lambda_N (2\pi)^D$ to absorb an overall coefficient in momentum integrals. The self-energy (or bubble) one-particle-irreducible diagram $\sim \cV^2 \int \tilde G^2$ is the one-loop correction to the two-point function:
\ba
\parbox{4.1cm}{\includegraphics[width=4cm]{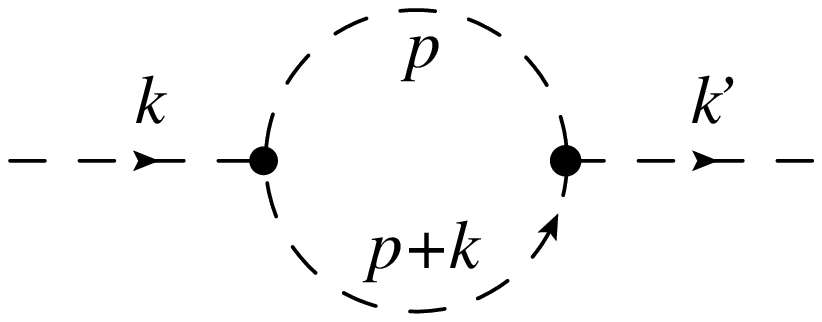}}  &=& \rmi \Pi(k,k')=\rmi\de^D(k-k')\tilde\Pi(k^2),\label{bubbd}
\ea
where the outer segments are external legs, not propagators, and
\ba
\hspace{-1cm}\tilde\Pi(k^2)&:=&\frac{\la_3^2}{2\rmi}\int\rmd^Dp\frac{1}{(p^2+m^2-\rmi\e)^\g[(k+p)^2+m^2-\rmi\e]^\g}\label{tip}\\
              &\stackrel{p^0\to \rmi p^0}{\rightarrow}& \frac{\la_3^2}{2}\int\rmd^Dp\,\frac{1}{(p^2+m^2)^\g[(k+p)^2+m^2]^\g}\nonumber\\
&=& \frac{\la_3^2}{2\Gamma^2(\g)}\int\rmd^Dp\int_0^{+\infty}\rmd\s_1\rmd\s_2\,(\s_1\s_2)^{\g-1}\rme^{-\s_1(p^2+m^2)-\s_2[(k+p)^2+m^2]}\nonumber\\
&=& \frac{\la_3^2}{2\Gamma^2(\g)}\int_0^{+\infty}\rmd\s_1\rmd\s_2\,(\s_1\s_2)^{\g-1}\rme^{-(\s_1+\s_2)m^2}\int\rmd^Dp\,\rme^{-\s_1p^2-\s_2(k+p)^2},\nonumber
\ea
where we analytically continued all momenta to Euclidean momentum space (therefore, this is not Efimov continuation since also external energies are imaginary). Using the Feynman pa\-ram\-e\-triza\-tion $x:=\s_1/y$ and $y:=\s_1+\s_2$ (so that $\s_1=xy$ and $\s_2=y(1-x)$), the last exponent in the above expression reads
\be\nonumber
\fl \s_1p^2+\s_2(k+p)^2=y[xp^2+(1-x)(k+p)^2]=y[p+(1-x)k]^2+yx(1-x)k^2\,,
\ee
and, calling ${p'}^\mu=p^\mu+(1-x) k^\mu$, we get
\ba
\fl \tilde\Pi(k^2) &=& \frac{\la_3^2}{2\Gamma^2(\g)}\int_0^1\rmd x\int_0^{+\infty}\rmd y\,y[y^2x(1-x)]^{\g-1}\rme^{-y [m^2+x(1-x)k^2]}\int\rmd^Dp'\,\rme^{-y {p'}^2}\nonumber\\
\fl &=& \frac{\la_3^2\pi^{\frac{D}{2}}}{2\Gamma^2(\g)}\int_0^1\rmd x\,[x(1-x)]^{\g-1}\int_0^{+\infty}\rmd y\,y^{2\g-1-\frac{D}{2}}\rme^{-y [m^2+x(1-x)k^2]}\nonumber\\
\fl &=&\frac{\la_3^2\pi^{\frac{D}{2}}}{2\Gamma^2(\g)}\Gamma\left(2\g-\frac{D}{2}\right)\int_0^1\rmd x\,[x(1-x)]^{\g-1}[x(1-x)k^2+m^2]^{\frac{D}{2}-2\g},\label{interme}
\ea
where we took into account the Jacobian determinant $y$ and the last step holds only for $2\g-D/2>0$ but it can be analytically continued to $2\g-D/2<0$ if the gamma function at the numerator does not diverge (we will come back to this important point later). One can verify that in the limit $\g\to 1$ and $D\to 4-2\ve$ one obtains, after subtracting the $1/\ve$ divergence, the standard bubble diagram in four-dimensional canonical scalar field theory \cite{Sch14}.

Calling $z:=4x(1-x)$,
\ba
\fl \tilde\Pi(k^2) &=&\frac{\la_3^2\pi^{\frac{D}{2}}}{2^{2\g}\Gamma^2(\g)}\Gamma\left(2\g-\frac{D}{2}\right)\int_0^1\rmd z\,(1-z)^{-\frac12}z^{\g-1}\left(\frac14zk^2+m^2\right)^{\frac{D}{2}-2\g}\nonumber\\
\fl &=& \frac{\la_3^2\pi^{\frac{D}{2}}}{2}\frac{\Gamma\left(2\g-\frac{D}{2}\right)}{\Gamma(2\g)} (m^2)^{\frac{D}{2}-2\g}{}_2F_1\left(\g,\,2\g-\frac{D}{2};\,\g+\frac12;\,-\frac{k^2}{4m^2}\right), \label{tip3}
\ea
valid for $\g>0$ and analytically continuable to $\g<0$. ${}_2F_1(a,b;c;z)=\sum_{n=0}^{+\infty}[(a)_n(b)_n/(c)_n] z^n/n!$ is the hypergeometric function, where $(a)_n=\Gamma(a+1)/\Gamma(a+1-n)$. The expression in Lorentzian momenta is obtained from \Eq{tip3} under the replacement $k^2\to\k^2-\rmi\e$, with $k^2$ in Lorentzian signature.

The self-energy \Eq{tip3} is finite provided $\Gamma(2\g-D/2)$ does not diverge. Thus, 
\be\label{rg3}
\boxd{\g\neq \frac{D}{4}-\frac{n}{2}\,,\qquad n\in\mathbb{N}\,.}
\ee
 In $D=4$ dimensions, 
\be\label{rg4}
\g\neq \frac{2-n}{2}= 1,\,\frac12,\,0,\,-\frac12,\dots\,,
\ee
where negative values are excluded by the positivity bound \Eq{posibo}. In particular, $\tilde\Pi$ diverges in the double limit $\g\to 1$, $D\to 4$, the standard four-dimensional theory.

For a massless theory, one can integrate \Eq{interme} with $m^2=0$ to get
\ba
\tilde\Pi(k^2)&=&\frac{\la_3^2\pi^{\frac{D+1}{2}}}{2^{D-2\g}}\frac{\Gamma\left(\frac{D}{2}-\g\right)\Gamma\left(2\g-\frac{D}{2}\right)}{\Gamma^2(\g)\Gamma\left(\frac12+\frac{D}{2}-\g\right)}\,(k^2)^{\frac{D}{2}-2\g}\nonumber\\
&=&\frac{\la_3^2\pi^{\frac{D}{2}}}{2}\frac{\Gamma^2\left(\frac{D}{2}-\g\right)\Gamma\left(2\g-\frac{D}{2}\right)}{\Gamma^2(\g)\Gamma(D-2\g)}\,(k^2)^{\frac{D}{2}-2\g}\,,\label{m0Pi}
\ea
if $\g<D/2$. This range already excludes the divergence points $\g=D/2+n$ coming from the first $\Gamma$ in the numerator, where $n\in\mathbb{N}$, so that the finiteness condition is again \Eq{rg3}.

\subsubsection{Dyson propagator and one-loop unitarity.}\label{dypro}

In this sub-section, we write the full quantum propagator $-\rmi \tilde G_{\rm Dyson}$ at all orders in perturbation theory as a Dyson series of one-particle-irreducible diagrams. Consider a diagram consisting in the one-loop self-energy \Eq{bubbd} to which a bare propagator in $k$ replaces the left external leg. Integrating over the internal momentum $k$ makes this a truncated vertex, to which one can attach any other diagram with outgoing momentum $k$:
\ba
\parbox{2.6cm}{\includegraphics[width=2.5cm]{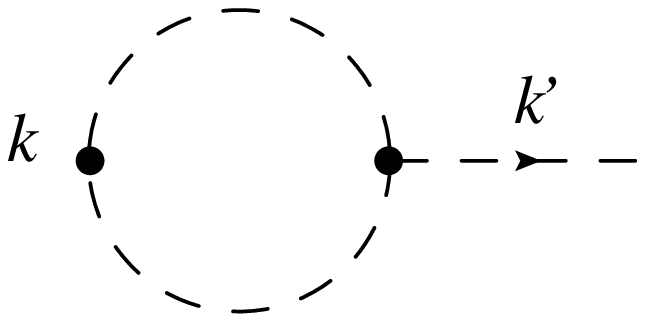}}  = \cB&=&\int\rmd^D k\,\frac{-\rmi}{(k^2+m^2-\rmi\e)^\g}\,\rmi\Pi(k,k')\nonumber\\
&=&\frac{1}{({k'}^2+m^2-\rmi\e)^\g}\,\tilde\Pi({k'}^2)=\tilde G\tilde\Pi\,.\label{bloc}
\ea
The iteration of this diagram gives the Dyson series for the full quantum Green's function:
\ba
\tilde G_{\rm Dyson} &=&\tilde G+\cB\tilde G+\cB(\cB\tilde G)+\dots=(1-\cB)^{-1}\tilde G=\frac{1}{\tilde G^{-1}-\tilde\Pi}\nonumber\\
&=&\frac{1}{(k^2+m^2-\rmi\e)^\g-\tilde\Pi(k^2)}\label{dys}\,.
\ea
Having already established a set of conditions to make $\tilde\Pi$ finite, we can extend the above conclusions on renormalizability to all loop levels. This expression does not include higher-loop irreducible diagrams and their level of divergence should be studied separately. However, there is a good chance that these diagrams do not change the picture because not only they are composed of finite bubble diagrams, buy they also have a larger number of external legs than one-particle-irreducible diagrams, which reduces the superficial degree of divergence. Therefore, we expect also these higher-loop irreducible diagrams to be finite.

From \Eq{rhoG} and \Eq{dys}, we can check one-loop unitarity, which we will do only in the massless case $m^2=0$. Assuming $\Im\,\tilde\Pi\neq 0$, we can neglect the $\rmi\e$ term in \Eq{dys} and write 
\be
\Im\,\tilde G_{\rm Dyson} = \frac{\Im\,\tilde\Pi(k^2)}{[(k^2+m^2)^\g-\Re\,\tilde\Pi(k^2)]^2+[\Im\,\tilde\Pi(k^2)]^2}\,,
\ee
so that we have that $\rho(s)\geq 0$ if, and only if,
\be\label{rhoG1}
\lim_{\e\to 0^+}\Im\,\tilde\Pi(-s-\rmi\e)\geq 0\,.
\ee
The calculation is the same as the one leading to \Eq{rhofinal} but with $m^2=0$, $\g$ replaced by $2\g-D/2$ and an overall coefficient that can be read off from \Eq{m0Pi}:
\ba
\rho(s)&\propto& \lim_{\e\to 0^+}\frac{1}{\pi}\Im\,\tilde\Pi(-s-\rmi\e)\nonumber\\
&=&\frac{\la_3^2\pi^{\frac{D}{2}}}{2}\frac{\Gamma^2\left(\frac{D}{2}-\g\right)\Gamma\left(2\g-\frac{D}{2}\right)}{\Gamma^2(\g)\Gamma(D-2\g)}\frac{\sin\left[\pi\left(2\g-\frac{D}{2}\right)\right]}{\pi}\frac{1}{s^{2\g-\frac{D}{2}}}\nonumber\\
&=&\frac{\la_3^2\pi^{\frac{D}{2}}}{2}\frac{\Gamma^2\left(\frac{D}{2}-\g\right)}{\Gamma^2(\g)}\frac{1}{\Gamma(D-2\g)\Gamma\left(1+\frac{D}{2}-2\g\right)}\frac{1}{s^{2\g-\frac{D}{2}}}\,.\label{rhofinalm0}
\ea
Recalling that \Eq{m0Pi} was calculated for $\g<D/2$, this expression is positive semi-definite if, and only if, $\Gamma(1+D/2-2\g)> 0$, i.e.,
\be\label{rg1l}
\fl \g<\frac{D+2}{4}\,,\qquad \frac{D+4}{4}+n<\g<\frac{D+6}{4}+n\,,\qquad n=0,1,\dots,\left\lfloor \frac{D-1}{2}\right\rfloor.
\ee
For example, in $D=4$ dimensions one gets $n=0,1$ and
\be
D=4:\qquad \g\in \left(-\infty,\frac{3}{2}\right)\cup\left(2,\frac{5}{2}\right)\cup\left(3,\frac{7}{2}\right)\,.
\ee

One could extend the unitarity range by analytically continue \Eq{m0Pi} to $\g>D/2$, but it is not necessary. In fact, in order for the perturbative theory to be well-defined the unitarity range should be the same at all orders. Comparing the free-level range \Eq{rg1} with \Eq{rg1l}, we see that their intersection coincides with the range \Eq{rg1}. Therefore, we take \Eq{rg1} as \emph{the} unitarity range of the theory and conjecture that the unitarity range at higher loops will always contain \Eq{rg1}. We recover these findings, and extend them to the case $m\neq 0$, with the reflection-positivity method of appendix \ref{appB}.

\subsubsection{Range of \texorpdfstring{$\g$}{gamma}.}\label{rangega1}

Comparing the bounds \Eq{rg2} coming from power-counting renormalizability and \Eq{rg1} from unitarity, one might be induced to conclude that it is not possible to make a fundamental quantum field theory with the fractional d'Alembertian because one encounters either infinitely many divergences in the UV or instabilities and negative-norm states.

However, we also showed that the power-counting argument does not tell the whole story and one can get unitarity and one-loop finiteness, and possibly even finiteness at all orders, if $\g$ falls in one of the ranges \Eq{rg1} and does not pick any of the values \Eq{rg4}. For example, if $\g$ is positive in order to have a positive spectral dimension according to \Eq{posibo}, then the four-dimensional fractional theory is one-loop finite in the intervals $0<\g<1/2$ and $1/2<\g<1$.


\subsection{Theory \texorpdfstring{$T[\B+\B^\g]$}{TBBg}}

\subsubsection{Kinetic term and action.}\label{kite2}

To get a multi-scale geometry with explicit scaling, we must combine the operator \Eq{Kgfix} with at least one length scale $\ell_*$. There are at least two ways in which one can do that.
\begin{itemize}
\item Combine operators with different masses as done by Trinchero \cite{Tri12,Tri17,Tri18}:
\be\label{trinky}
\cK=\cK_{\rm T}:=(\B-m^2)(E_*^2-\B)^{\g-1},
\ee
where $m$ is an arbitrary mass scale that can be much smaller than the characteristic energy $E_*=\ep\lp/\ell_*$. The IR limit (momentum-energy scales $\ll E_*$) of this operator is
\be
\cK_{\rm T}\simeq E_*^{2(\g-1)}(\B-m^2)\,,\qquad k\ll E_*\,,
\ee
and redefining the scalar as $\phi_{\rm eff}=E_*^{\g-1}\phi$, one gets the standard scalar field theory with ordinary dimensionality $[\phi_{\rm eff}]=(D-2)/2$. In the UV limit (momentum-energy scales $\gg E_*$),
\be
\cK_{\rm T}\simeq (-\B)^\g\,,\qquad k\gg E_*\,,
\ee
and the theory becomes effectively massless. 
\item Take a sum of fractional d'Alembertians with the same mass, in line with the paradigm of multi-fractal geometries where scale-dependent correlation functions are described by a discrete set of critical exponents \cite{frc4,revmu}. Dimensional flow is realized here by explicit multi-scaling \cite{mf0}, i.e., the fundamental scales of the geometry appear directly in the coefficients of the sum of operators. The minimum to get a non-trivial dimensional flow is one fundamental scale and two exponents, 1 and $\g\neq 1$. Therefore, we propose
\be\label{Kg}
\cK= \ell_*^{2(1-\g)}(\B-m^2)-(m^2-\B)^\g\,,
\ee
where one could naturally identify $\ell_*=m^{-1}$ (we do not do it in order to allow for the massless case). With this definition, the propagator of the free theory diverges at $-k^2=m^2$ at all scales. Note that \Eq{Kg} is similar to the definition of distributed-order fractional derivatives ${\rm D}:=\int_0^\g\rmd\g'\,\mu(\g')\,\p^{\g'}$ \cite{Cap69,Cap95,BT1,BT2,CGS,LoH02,Koc1,Koc2}, where one integrates over a parameter $\g'$ with a weight $\mu(\g')$. If $\mu(\g')\simeq \ell_*^{2(1-\g)}\de(\g'-1)-\de(\g'-\g)$, we get \Eq{Kg}. To get a realistic dimensional flow, it is not necessary to consider a more complicated $\mu(\g')$.
\end{itemize}
Plugging the kinetic term \Eq{Kg} into the action \Eq{actss} with $v=1$, the action for the theory $T[\B+\B^\g]$ is
\be\label{actssfin}
\fl \boxd{S=\int\rmd^Dx\,\left\{\frac12\phi\left[\ell_*^{2(1-\g)}(\B-m^2)-(m^2-\B)^\g\right]\phi-V(\phi)\right\},}
\ee
where, again, the potential includes only non-linear interactions.

The equation of motion $\de S/\de\phi=0$ from the action \Eq{actssfin} is calculated using the integral representation \Eq{intpar} and reads
\be\label{scaom}
\left[\ell_*^{2(1-\g)}(\B-m^2)-(m^2-\B)^\g\right]\phi-V'(\phi)=0\,.
\ee
Free-field solutions can be found according to the general scheme presented in section \ref{freeeq} and the final result therein is valid also for this theory in the UV limit. The propagator in the UV is the one calculated in section \ref{secpro}.

\subsubsection{Dimensional flow.}\label{dimflo2}

In the theory $T[\B+\B^\g]$ with the measure choice \Eq{v1}, dimensional flow is the same described in section \ref{dimflo20}. The Hausdorff dimension is constant, equation \Eq{dhgene}, while the spectral dimension varies according to one of the following alternatives.
\begin{itemize}
\item When $\g>1$, the fractional operator in \Eq{Kg} dominates the dynamics in the UV, equation \Eq{dsuv}, and the renormalization calculations of section \ref{sec4g} apply to the UV limit of the theory.
\item When $\g<1$, the spectral dimension is anomalous in the IR and standard in the UV, equation \Eq{dsir}, and renormalizability is not improved.
\item Extending the operator \Eq{Kg} to three operators with three different exponents and two scales $\ell_1<\ell_2$,
\be\label{Kg3}
\fl \cK= (\B-m^2)-\ell_1^{2(\g_1-1)}(m^2-\B)^{\g_1}-\ell_2^{2(\g_2-1)}(m^2-\B)^{\g_2}\,,\qquad \g_2<1<\g_1\,,
\ee
we get the three-regime dimensional flow of \Eq{ds3reg}.
\end{itemize}

\subsubsection{Unitarity, renormalization and range of \texorpdfstring{$\g$}{gamma}.}\label{rangega2}

Considering the multi-fractional extension of the Green's function \Eq{Gfrac},
\be\label{Gmfrac}
\tilde G(-k^2) = \frac{1}{\ell_*^{2(1-\g)}(k^2+m^2-\rmi\e)+(k^2+m^2-\rmi\e)^\g}\,,
\ee
with the method discussed in sections \ref{kalere} and \ref{unifrac} one can show that
\be
\fl \rho(s)=\frac{1}{\pi}\frac{\sin(\pi\g)}{\ell_*^{4(1-\g)} (s-m^2)^{2-\g}+(s-m^2)^\g-2 \ell_*^{2(1-\g)}\cos(\pi\g)(s-m^2)}\,,
\ee
which is positive definite in the range \Eq{rg1}, in particular in the interval $0<\g<1$ \Eq{rg1min}, and it reproduces \Eq{rhofinal} in the UV regime $\ell_*k\ll 1$. In the IR regime $\ell_*k\gg 1$, one does not recover the delta \Eq{poi} unless one takes the limit $\e\to 0^+$ after approximating the expression to the Poisson kernel.

Concerning one-loop renormalization, since we are interested in the fate of UV divergences, we can consider only the integer or the fractional part of the multi-scale propagator \Eq{Gmfrac}, depending on whether $\g<1$ or $\g>1$, respectively. The integer sector is a standard QFT, while the fractional sector was studied in sections \ref{1lreno1}--\ref{dypro}.

The conclusions laid out in section \ref{rangega1} hold for a theory with a single fractional operator as a kinetic term. However, in the multi-fractional case \Eq{actssfin} having $\g<1$ means that the fractional limit we studied in sections \ref{1lreno1}--\ref{dypro} corresponds to the IR limit of the theory, not the UV one. Therefore, the multi-fractional theory \Eq{actssfin} can never be one-loop finite and unitary at the same time. 


\subsection{Theory \texorpdfstring{$T[\B^{\g(\ell)}]$}{TBgell}}

\subsubsection{Kinetic term and action.}\label{kite3}

In this version of the multi-scale dynamics with fractional d'Alembertian, one can make the parameter $\g$ scale dependent \cite{fra6,frc4} or coordinate dependent \cite{Sam13}. In the first case, which is simpler, the kinetic operator becomes
\be\label{Kell}
\cK=\cK_{\g(\ell)}=-(m^2-\B)^{\g(\ell)}\,,
\ee
where we can choose the profile $\g(\ell)$ given in \Eq{Kellg}. When computing the propagator or Feynman diagrams, one may find the Green function of \Eq{Kell} much simpler to handle than that of \Eq{Kg}.

The action gets an overall extra integration over the probed scale as in \Eq{actvog},
\be\label{actgell}
\boxd{S=\frac{1}{\ell_*}\int_0^{+\infty}\rmd\ell\,\tau(\ell)\int\rmd^Dx\left[-\frac12\phi(m^2-\B)^{\g(\ell)}\phi-V(\phi)\right],}
\ee
where $\tau(\ell)$ is a one-parameter weight. The equation of motion is
\be\label{scaomgl}
(m^2-\B)^{\g(\ell)}\phi+V'(\phi)=0\,.
\ee
Its solutions in the free case and the propagator were discussed in, respectively, sections \ref{freeeq} and \ref{secpro}, where we have to replace $\g\to\g(\ell)$.

\subsubsection{Dimensional flow.}\label{dimflo3}

This is the same flow as in $T[\p^{\g(\ell)}]$:
\begin{itemize}
\item With the profile \Eq{Kellg} for any positive $\g$, one gets the dimensional flow \Eq{dsuv}.
\item With the profile \Eq{newprof} for any positive $\g$, one recovers \Eq{dsir}.
\item With the profile \Eq{prof2} for any positive $\g_1$ and $\g_2$, we get the flow \Eq{ds3reg}: a UV regime where renormalizability can improve, a mesoscopic one where standard classical and quantum field theory is recovered, and an IR or ultra-IR one which may be relevant for cosmology.
\end{itemize}

\subsubsection{Unitarity, renormalization and range of \texorpdfstring{$\g$}{gamma}.}\label{rangega3}

The calculation done in section \ref{unifrac} holds also for $T[\B^{\g(\ell)}]$. The theory is unitary if $\g$ falls within one of the bounds \Eq{rg1} at all scales $\ell$.

The results of sections \ref{1lreno1}--\ref{dypro} hold, since in any Feynman diagram all exponents $\g(\ell)$ coming from different internal lines are evaluated at the same scale $\ell$.

In the theory \Eq{actgell}, the limit of fractional exponent $\g(\ell)\to\g$ is in the UV by construction assuming \Eq{Kellg}, even when $\g<1$. Therefore, we can export to $T[\B^{\g(\ell)}]$ the results of section \ref{rangega1} on the range of $\g$ preserving unitarity and renormalizability. 


\section{Comparison with the literature}\label{actimm}

In this section, we state what of the above is new and what was known from the literature on the scalar field, recapitulating past results to the best of our knowledge.
\begin{itemize}
\item The mathematical properties of the fractional d'Alembertian $\cK(\B)=(-\B)^\g$ acting on a scalar field have been studied extensively \cite{BGG,Mar91,Gia91,BGO,BG,doA92,Mar93,BOR,BBOR1,Barci:1996ny,BBOR2}. Using Caffarelli--Silvestre extension theorem \cite{CaSi}, it was recently shown that the $D$-dimensional action of the massless free scalar field $\phi(x)$ with $\cK(\B)=(-\B)^\g$ is equivalent to the $(D+1)$-dimensional action of a scalar $\Phi(x,y)$ with an extra fictitious spatial direction with fractional unilateral measure \cite{Frassino:2019yip}:
\ba
S&=&\frac12\int\rmd^Dx\,\phi(x)\,(-\B)^\g\phi(x)\nonumber\\
&=&-\frac{2^{2(\g-1)}\G(\g)}{\g\G(-\g)}\int\rmd^Dx\int_0^{+\infty}\rmd y\,y^{1-2\g}\,\p_M\Phi(x,y)\,\p^M\Phi(x,y),\label{corres}
\ea
where $M=(\mu,y)=0,1,\dots,D-1,D$ and $\lim_{y\to 0}y^{1-2\g}\p_y\Phi(x,y)=0$. This correspondence allows for a smooth quantization and expands to the realm of fractional operators the notion, valid for non-local quantum gravity with exponential or asymptotically polynomial operators \cite{Calcagni:2018lyd,Calcagni:2018gke}, that it is possible to recast non-local systems as higher-dimensional local systems. The infinite number of initial conditions of non-local dynamics translate into field boundary conditions along the extra direction. However, \Eq{corres} is surprising also for another reason, unnoticed in \cite{Frassino:2019yip}: modulo the overall constant, the last line of \Eq{corres} coincides with the action of a free massless scalar field in the multi-fractional theory $T_1$ with normal derivatives and unilateral fractional measure in the $y$ direction \cite{fra1,fra2,fra3,frc2}. This suggests a relation between $T[\B^\g]$ and $T_1$ we will comment upon in the conclusions.
\item The canonical quantization of a free scalar theory with $\cK(\B)=\B(-\B)^{-\a}$ with $0<\a<1$ has been carried out in \cite{doA92,BOR}.
\item Trinchero \cite{Tri12,Tri17,Tri18} considered the unitarity and renormalization properties of a Euclidean theory with operator \Eq{trinky},
\be\label{trinky2}
\cL=\frac12\phi(\B-m^2)(E_*^2-\B)^{-2\a}\phi-\la\phi^4\,,\qquad \a=\frac{1-\g}{2}
\ee
with $m\neq E_*$ and $\a>0$ ($\g<1$). 
\item A non-local kinetic term $\cK=(\B-m^2)\sqrt{E_*^2-\B}$ can arise for a scalar field in $\k$-Minkowski non-commutative spacetime \cite{KoWa}. The propagator and its branch cuts were studied in \cite{ACo}.
\item The canonical quantization of a free scalar was generalized to an arbitrary $\cK(\B)$ with branch cuts in \cite{BOR} and later in \cite{Belenchia:2014fda,Saravani:2015rva,Belenchia:2016sym,Belenchia:2017yvv,Saravani:2018rwm}. In \cite{Barci:1996ny} (reviewed in \cite{Belenchia:2014fda}), it was shown that only massless states appear asymptotically in the free quantum theory and, if only branch cuts are present, then the only asymptotic state is the vacuum. The Huygens' principle for this class of models was discussed in \cite{Gia91,BG,Belenchia:2014fda,Belenchia:2017yvv}. A class of operators $\cK(\B)$ gives rise to a model which does not admit a variational principle and such that the modes on the branch cut cannot be detected through scattering experiments, since they never appear as in-states in non-zero amplitudes \cite{Saravani:2015rva}. Since these modes do not interact, they can serve as a dark-matter candidate \cite{Saravani:2015rva}. The interpretation of the continuum of modes of the branch cut as infinitely many local scalars has been studied in \cite{Saravani:2018rwm}.
\item A non-minimally coupled scalar field with fractional Laplacian $\Delta^{3/2}$ was introduced to preserve detailed balance in Ho\v{r}ava--Lifshitz gravity in the matter sector \cite{Calcagni:2009qw}.
\end{itemize}
Our scalar theory differs from the above proposals not only in the form of the kinetic operator but also in the motivation and the focus. Regarding the justification, some previous proposals share an interest in toy models of quantum gravity, but while \cite{Tri12,Tri17,Tri18} were motivated by non-commutative geometry and \cite{Belenchia:2014fda,Saravani:2015rva,Belenchia:2016sym,Belenchia:2017yvv,Saravani:2018rwm} by causal sets, our action \Eq{actssfin} has been built from basic considerations of multi-fractal geometry \cite{revmu,frc4,mf0}. This change in perspective also accounts for the different stress in the geometrical interpretation of the theory, here more centered on dimensional flow. Moreover, our focus is on unitarity and renormalizability, while previous works mainly studied the canonical quantization and the unitarity of their models, except \cite{Tri12,Tri18} where some Feynman diagrams were calculated for the theory \Eq{trinky2}.

Causal sets, non-commutative geometry and multi-fractal geometry are all connected within the bigger scheme of quantum gravity \cite{revmu,CaRo1}, which explains how independent reasonings led to operators belonging to the same mathematical class of fractional derivatives. The fractional proposals of \cite{Tri12,Tri17,Tri18,Belenchia:2014fda,Saravani:2015rva,Belenchia:2016sym,Belenchia:2017yvv,Saravani:2018rwm} appeared after the multi-fractal and multi-fractional theories, where fractional operators were invoked in quantum gravity as early as \cite{fra2,fra4,frc1}, but all of them were influenced by previous studies on the fractional d'Alembertian. 


\section{Conclusions}\label{conc}

In this paper, we studied the classical and quantum properties of scalar theories with fractional kinetic terms, respecting or violating Lorentz invariance. 

The theories with fractional derivatives, labelled $T[\p^\g]$, $T[\p+\p^\g]$ and $T[\p^{\g(\ell)}]$ do not have Lorentz symmetry and they are technically difficult due to the presence of fractional derivatives. Here we have defined basic aspects of their classical and quantum dynamics.

The theories labelled $T[\B^\g]$, $T[\B+\B^\g]$ and $T[\B^{\g(\ell)}]$ are Lorentz invariant and are those that we worked out more extensively. We showed that, in general, it is difficult to choose a value of $\g$ accommodating both unitarity and renormalizability. When the kinetic operator is a single fractional d'Alembertian, the theory $T[\B^\g]$ can be made unitary and one-loop finite, with evidence that it may be finite at all loops.

Our understanding of these scalar theories is incomplete. For instance, the propagation of signals is an open problem. The Huygens principle states that the Green's function has support on the light cone, i.e., signals propagate with the speed of light. Results with non-local operators with branch cuts indicate that the Huygens principle is violated for certain powers of the d'Alembertian and that virtual-particle propagation happens also inside the light cone \cite{BG,Belenchia:2014fda}, as in \Eq{Gretad}. Whether this affects also propagation of physical signals remains to be seen.

Another question is whether the results found here can be extended to the gravitational sector and are representative of the features of quantum gravity with fractional operators. We tackle this problem in \cite{mf2}.


\section*{Acknowledgments}

The author is supported by the I+D grants FIS2017-86497-C2-2-P and PID2020-118159GB-C41 of the Spanish Ministry of Science and Innovation. He thanks F Briscese, L Modesto, L Rachwa\l\ and especially G Nardelli for useful comments.



\begin{appendices}


\section{Properties of fractional derivatives}\label{appA}

The Liouville derivative \Eq{lio},
\be\label{lioapp}
\fl {}_\infty\p^\g f(x) := \frac{1}{\Gamma(m-\g)}\int_{-\infty}^{x}\, \frac{\rmd x'}{(x-x')^{\g+1-m}}\p_{x'}^m f(x')\,,\qquad m-1\leq \g<m\,,
\ee
obeys the following properties.
\begin{itemize}
\item[(a)] Limit to ordinary calculus:
\be\label{limlio}
\lim_{\g\to n}{}_\infty\p^\g = \p^n\,,\qquad \g=n\in\mathbb{N}.
\ee
\item[(b)] Linearity:
\be\label{linlio}
{}_\infty\p^\g[c_1 f(x)+c_2 g(x)]=c_1 ({}_\infty\p^\g f)(x)+c_2({}_\infty\p^\g g)(x)\,.
\ee
\item[(c)] Commutation:
\be\label{comlio}
{}_{\infty}\p^\g\,{}_{\infty}\p^\b = {}_{\infty}\p^\b\,{}_{\infty}\p^\g={}_{\infty}\p^{\g+\b}\,,\qquad \forall~ \g,\b>0\,.
\ee
\item[(d)] Kernel:
\be
\fl {}_{\infty}\p^\g x^\b = (-1)^{\g}\frac{\Gamma(\b+1)}{\Gamma(\b+1-\g)}\frac{\sin(\pi\b)}{\sin[\pi(\b-\g)]} x^{\b-\g}=(-1)^{\g}\frac{\Gamma(\g-\b)}{\Gamma(-\b)} x^{\b-\g}.\label{pllio}
\ee
Equation \Eq{pllio} vanishes for $\b=0,1,2,\ldots,m-1$ and is ill-defined for $\b=\g$. In particular, the Liouville fractional derivative of a constant is zero and the kernel of a Liouville derivative with $0<\g<1$ is trivial:
\be\label{1lio}
{}_{\infty}\p^\g 1=0\,,\qquad 0<\g<1\,.
\ee
\item[(e)] Eigenfunctions:
\be
{}_{\infty}\p^\g \rme^{\la x} = \la^\g\rme^{\la x}\,.\label{explio}
\ee
\item[(f)] Leibniz rule:
\be\label{leru}
{}_\infty\p^\g(fg)=\sum_{j=0}^{+\infty}\frac{\Gamma(1+\g)}{\Gamma(\g-j+1)\Gamma(j+1)} (\p^j f)({}_\infty\p^{\g-j}g)\,.
\ee
\item[(g)] Integration by parts:
\be\label{ibp}
\int_{-\infty}^{+\infty}\rmd x\, f\,{}_{\infty}\p^\g g = \int_{-\infty}^{+\infty}\rmd x\, ({}_{\infty}\bp^\g f)\,g\,.
\ee
\end{itemize}
With minor changes, the above formul\ae\ hold also for the Weyl derivative \Eq{wey},
\be\label{weyapp}
\fl {}_\infty\bar\p^\g f(x) := \frac{1}{\Gamma(m-\g)}\int_{x}^{+\infty}\, \frac{\rmd x'}{(x'-x)^{\g+1-m}}\p_{x'}^m f(x')\,,\qquad m-1\leq \g<m\,.
\ee
\begin{itemize}
\item[(a)] Limit to ordinary calculus:
\be\label{limwey}
\lim_{\g\to n}{}_\infty\bp^\g = (-1)^n\p^n\,,\qquad \g=n\in\mathbb{N}.
\ee
\item[(b)] Linearity:
\be\label{linwey}
{}_\infty\bp^\g[c_1 f(x)+c_2 g(x)]=c_1 ({}_\infty\bp^\g f)(x)+c_2({}_\infty\bp^\g g)(x)\,.
\ee
\item[(c)] Commutation:
\be\label{comwey}
{}_{\infty}\bp^\g\,{}_{\infty}\bp^\b = {}_{\infty}\bp^\b\,{}_{\infty}\bp^\g={}_{\infty}\bp^{\g+\b}\,,\qquad \forall~ \g,\b>0\,.
\ee
\item[(d)] Kernel:
\be
{}_{\infty}\bp^\g x^\b =\frac{\Gamma(\b+1)}{\Gamma(\b+1-\g)}\frac{\sin(\pi\b)}{\sin[\pi(\b-\g)]} x^{\b-\g}=\frac{\Gamma(\g-\b)}{\Gamma(-\b)} x^{\b-\g}.\label{plwey}
\ee
Equation \Eq{plwey} vanishes for $\b=0,1,2,\ldots,m-1$ and is ill-defined for $\b=\g$. In particular, the Weyl fractional derivative of a constant is zero and the kernel of a Weyl derivative with $0<\g<1$ is trivial:
\be\label{1wey}
{}_{\infty}\bp^\g 1=0\,.
\ee
\item[(e)] Eigenfunctions:
\be
{}_{\infty}\bp^\g \rme^{\la x} = (-\la)^\g\rme^{\la x}\,.\label{expwey}
\ee
\item[(f)] Leibniz rule:
\be\label{lerub}
{}_\infty\bp^\g(fg)=\sum_{j=0}^{+\infty}\frac{\Gamma(1+\g)}{\Gamma(\g-j+1)\Gamma(j+1)} (\p^j f)({}_\infty\bp^{\g-j}g)\,.
\ee
\item[(g)] Integration by parts: equation \Eq{ibp}.
\end{itemize}
The reader can find the proofs of all these statements in \cite{MR,Pod99,SKM,KST,frc1}.


\section{Alternative proof of unitarity}\label{appB} 

In this appendix, we check unitarity of the real scalar field theory $T[\B^\g]$ of section \ref{sec4g} using a different route than the spectral decomposition of section \ref{kalere}. Namely, we will show that the scalar product of field functionals in the Euclidean version of the theory is positive definite only for the set of ranges \Eq{rg1}. Our calculation is similar to that in \cite{Tri17} for the theory \Eq{trinky2} with kinetic operator \Eq{trinky}.

\begin{figure}
\bc
\includegraphics[width=12cm]{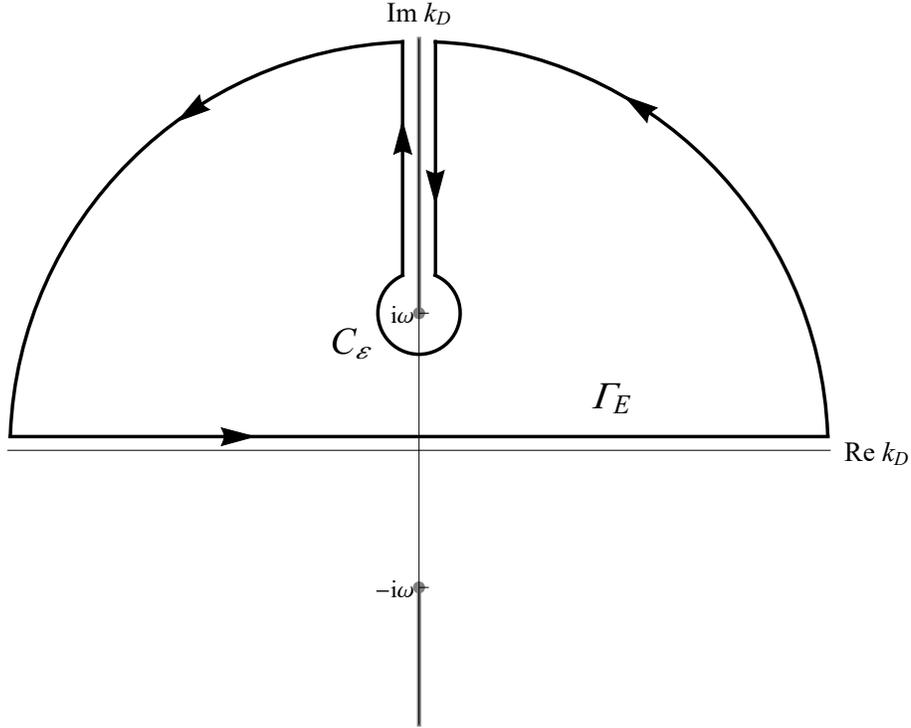}
\ec
\caption{\label{fig5} Contour $\Gamma_{\rm E}$ (black thick curve) in the $(\Re\, k_D,\Im\, k_D)$ plane.}
\end{figure}

We work in Euclidean position space with coordinates $x_1,x_2,\dots, x_D$ and define a reflection operation ${\rm R}$ with respect to the $x_D=0$ plane: ${\rm R} x_\mu = (-1)^{\de_{\mu D}}x_\mu$, i.e., ${\rm R}\bm{x}=\bm{x}$ for $\mu=1,\dots,D-1$ and ${\rm R} x_D=-x_D$. Let
\be\nonumber
\cF[\phi]=\int\rmd^Dx\,\vp(x)\,\phi(x)
\ee
be a linear functional of the field $\phi$, where $\vp$ is a test function with support in $x_D>0$. We denote the complex conjugate and reflected functional as $\cF_{\rm R}^*[\phi]=\int\rmd^Dx\,\vp^*({\rm R}x)\,\phi(x)$. Reflection positivity states that the expectation value of $\cF_{\rm R}[\phi]\cF[\phi]$  defined through the path integral is positive semi-definite:
\be\nonumber
\langle \cF_{\rm R}[\phi]\cF[\phi]\rangle:=\frac{1}{\int{\rm D}\phi\,\rme^{-S[\phi]}}\int{\rm D}\phi\,\rme^{-S[\phi]}\cF_{\rm R}^*[\phi]\cF[\phi]\geq 0\,,
\ee
where ${\rm D}\phi$ is the functional measure. This expectation value can be calculated as in standard QFT by introducing a current $\cJ\phi$ in the action and taking the second-order functional derivative $\de/\de\cJ$. The result is
\be
\langle \cF_{\rm R}[\phi]\cF[\phi]\rangle=\int\rmd^Dx\,\rmd^Dy\,\vp^*(x)\,G({\rm R}x-y)\,\vp(y)=:({\rm R}\vp,\vp)\,,
\ee
where $G$ is the Green's function, which in our case is
\be\label{Geuc}
G(x)=\int\frac{\rmd^D k}{(2\pi)^D}\,\frac{\rme^{-\rmi k\cdot x}}{(k^2+m^2)^\g}\,,
\ee
where $k^2=\sum_{\mu=1}^D k_\mu^2$. The test functions $\vp$ are by construction square integrable in this scalar product. For the sake of the calculation, one can take a `charge' distribution
\be\nonumber
\vp(x)=\sum_{i=1}^N q_i\de^D[x-x^{(i)}]\,,\qquad q_i\in\mathbb{C}\,.
\ee
Therefore, calling $\a_{\bm{k}}[\bm{x}^{(i)}]:=q_i\exp[\rmi \bm{k}\cdot\bm{x}^{(i)}]$ and $r_{ij}:=x_D^{(i)}+y_D^{(j)}$ and using definition \Eq{defom},
\ba
({\rm R}\vp,\vp) &=& \sum_{i,j}q_i^*q_j\int\rmd^Dx\,\rmd^Dy\,\de^D[x-x^{(i)}]\,G({\rm R}x-y)\,\de^D[y-y^{(j)}]\nonumber\\
					 &=& \sum_{i,j}q_i^*q_j G[{\rm R}x^{(i)}-y^{(j)}]\nonumber\\
					 &=&\sum_{i,j}q_i^*q_j\int\frac{\rmd^D k}{(2\pi)^D}\,\frac{\rme^{-\rmi k\cdot [{\rm R}x^{(i)}-y^{(j)}]}}{(k^2+m^2)^\g}\nonumber\\
					 &=&\sum_{i,j}\int_{-\infty}^{+\infty}\frac{\rmd^{D-1}\bm{k}}{(2\pi)^{D-1}}\a^*_{\bm{k}}[\bm{x}^{(i)}]\,\a_{\bm{k}}[\bm{y}^{(j)}]\int_{-\infty}^{+\infty}\frac{\rmd k_D}{2\pi}\,\frac{\rme^{\rmi k_Dr_{ij}}}{(k_D^2+\om^2)^\g}\nonumber\\
					 &=:& \sum_{i,j}\int\frac{\rmd^{D-1}\bm{k}}{(2\pi)^{D-1}}\a^*_{\bm{k}}[\bm{x}^{(i)}]\,I_\mathbb{R}^{ij}\,\a_{\bm{k}}[\bm{y}^{(j)}]\,,
\ea
where $\om^2=|\bm{k}|^2+m^2$, so that
\be\label{proofin}
({\rm R}\vp,\vp) \geq 0\qquad \Longleftrightarrow\qquad I_\mathbb{R}^{ij}\geq 0\,.
\ee
At this point, we analytically continue the Euclidean momentum $k_D$ to a complex momentum and perform the above integration making use of a contour $\G_{\rm E}$ in the $(\Re\, k_D,\Im\, k_D)$ plane, where ${\rm E}$ stands for Euclidean. The integrand has branch points on the imaginary axis at $k_D=\pm\rmi\om$ and branch cuts ramifying from those points to $\pm\rmi\infty$, respectively. Since $r_{ij}>0$, to get an exponential suppression $\exp(-\Im\, k_Dr_{ij})$ we choose $\G_{\rm E}$ in the $\Im\, k_D>0$ half plane: it runs along the real axis from $-\infty$ to $+\infty$ and makes a quarter arc counter-clockwise up to the branch cut $(\rmi\om,\rmi\infty)$, where the contour follows the cut down to the branch point and up again in the fourth quadrant, where it closes with another quarter arc counter-clockwise from the imaginary positive semi-axis to the negative real semi-axis (figure \ref{fig5}).

Since the integrand does not have any singularity inside the contour $\G_{\rm E}$, integration over $\G_{\rm E}$ gives zero. The contributions of the arcs at infinity is also zero thanks to the exponential suppression, so that
\be\nonumber
0=\int_{\G_{\rm E}}=\int_\mathbb{R}+\int_{C_\ve}+\int_{\rm cut}+\int_{\rm arcs}=\int_\mathbb{R}+\int_{C_\ve}+\int_{\rm cut},
\ee
implying
\be\nonumber
\int_\mathbb{R}=-\int_{C_\ve}-\int_{\rm cut}=-\int_{C_\ve}-\left(\int_{\rmi\infty+\ve}^{\rmi\om+\ve}+\int_{\rmi\om-\ve}^{\rmi\infty-\ve}\right).
\ee
Integration on the circle $C_\ve$ surrounding the branch point $\rmi\om$ can be done by a change of variable $k_D-\rmi\om=\ve\exp(\rmi\theta)$, where $-3\pi/2<\theta<\pi/2$:
\ba
\fl -\int_{C_\ve}\frac{\rmd k_D}{2\pi}\,\frac{\rme^{\rmi k_Dr_{ij}}}{(k_D^2+\om^2)^\g} &=&-\rmi\ve\int_{\frac{\pi}{2}}^{-\frac{3\pi}{2}}\frac{\rmd \theta\,\rme^{\rmi\theta}}{2\pi}\,\frac{\rme^{\rmi (\ve\rme^{\rmi\theta}+\rmi\om)r_{ij}}}{[\ve\rme^{\rmi\theta}(2\rmi\om+\ve\rme^{\rmi\theta})]^\g}\nonumber\\
\fl &=& \ve^{1-\g}\frac{\rme^{-\om r_{ij}}}{2\pi\rmi(2\rmi\om)^\g}\int_{\frac{\pi}{2}}^{-\frac{3\pi}{2}}\rmd \theta\,\rme^{\rmi\theta(1-\g)}+O(\ve^{2-\g})\nonumber\\
\fl &=& \ve^{1-\g}\frac{\rme^{-\om r_{ij}}}{\pi(1-\g)(2\om)^\g}\sin(\pi\g)+O(\ve^{2-\g})\,.\label{circle}
\ea
In the limit $\ve\to 0$, this contribution diverges for $\g>1$, is finite when $\g=1$ (ordinary case, no branch cut) and it vanishes for $\g<1$. Barring the standard case, we are forced to choose $\g<1$.

Having established that integration around the branch point gives zero, we are left to calculate the contribution of the branch cut. Changing variable into $k_D=\rmi\rho=\rme^{\rmi\pi/2}\rho$ in the first quadrant and $k_D=\rmi\rho'=\rme^{-3\rmi\pi/2}\rho'$ in the second quadrant, integration along the cut yields
\ba
I_\mathbb{R}^{ij}&=&-\int_{\rm cut}\frac{\rmd k_D}{2\pi}\,\frac{\rme^{\rmi k_Dr_{ij}}}{(k_D^2+\om^2)^\g}\nonumber\\
&=& -\rmi\int_{\infty}^{\om}\frac{\rmd\rho}{2\pi}\,\frac{\rme^{-\rho r_{ij}}}{(\rme^{\rmi\pi}\rho^2+\om^2)^\g}-\rmi\int_{\om}^{\infty}\frac{\rmd \rho'}{2\pi}\,\frac{\rme^{-\rho'r_{ij}}}{(\rme^{-\rmi\pi}\rho^{'2}+\om^2)^\g}\nonumber\\
&=& -\rmi\int_{\om}^{\infty}\frac{\rmd \rho}{2\pi}\,\rme^{-\rho r_{ij}}\frac{\rme^{\rmi\pi\g}-\rme^{-\rmi\pi\g}}{(\rho^2-\om^2)^\g}\nonumber\\
&=& \frac{\sin(\pi\g)}{\pi}\int_{\om}^{\infty}\rmd \rho\,\frac{\rme^{-\rho r_{ij}}}{(\rho^2-\om^2)^\g}\,.\label{finap}
\ea
The integrand is positive on a positive range, so that this expression is positive semi-definite if, and only if, $\sin(\pi\g)\geq 0$, i.e., when $2n\leq \g\leq 2n+1$ with $n\in\mathbb{Z}$. Since $\g<1$, the only ranges guaranteeing unitarity except the standard case are $0\leq\g< 1$, $-2\leq\g\leq-1$, $-3\leq\g\leq -4$, and so on, that is, the range \Eq{rg1}.

The same result can be obtained by performing the integral in \Eq{finap} explicitly using formula 8.432.3 of \cite{GR}, valid for $\g<1$:
\be\label{irij}
I_\mathbb{R}^{ij} = \frac{1}{\sqrt{\pi}\Gamma(\g)}\left(\frac{2\om}{r_{ij}}\right)^{\frac12-\g} K_{\frac12-\g}(\om r_{ij})\,,
\ee
where $K_{1/2-\g}$ is the modified Bessel function of the second kind, which is always positive. The sign of $I_\mathbb{R}^{ij}$ is thus determined by the one of $\Gamma(\g)$, which is positive when $\g>0$ or $-2n\leq \g\leq -2n+1$ with $n\in\mathbb{N}$ and $\g\neq 1$. This is exactly the same condition obtained from semi-positivity of the spectral function, equation \Eq{rhos}.

In the presence of interactions, the unitarity condition does not change, at least for a cubic potential at one-loop level. One can repeat the calculation of this appendix after replacing the bare Green's function \Eq{Geuc} with the Dyson Green's function \Eq{dys}, where $\tilde\Pi(k^2)$ was calculated in section \ref{1lreno2} for the $\phi^3$ theory. To leading order in a perturbative expansion, one can approximate the dispersion relation as
\ba
0&=&(k^2+m^2)^\g-\tilde\Pi(k^2)=(k^2+m^2)^\g-\sum_{n=0}^{+\infty}\frac{1}{n!}\tilde\Pi^{(n)}(-m^2)\,(k^2+m^2)^n\nonumber\\
&\simeq& (k_D^2+\om^2)^\g-\tilde\Pi(-m^2)\,,
\ea
provided $\g<1$. In the massless case, since $\tilde\Pi(0)=0$ there is no change in the spectrum of the theory to leading order and one-loop unitarity is preserved for the same values of $\g$ determined above. 

The $m\neq 0$ case is less trivial. If one had $\g=1$, one would have experienced the well-known pole displacement $k_D=\pm\rmi\om\to\pm\rmi\om\sqrt{1-\tilde\Pi(-m^2)/\om^2}\simeq \pm\rmi\om[1-\tilde\Pi(-m^2)/(2\om^2)]$ of the ordinary theory. However, when $\g$ is non-integer what happens is that the branch points $k_D=\pm\rmi\om$ remain in their place, while simple poles appear at 
\be
k_D=\pm\rmi\tilde\om:=\pm\rmi\om\sqrt{1-\frac{\tilde\Pi^\frac{1}{\g}(-m^2)}{\om^2}}\simeq \pm\rmi\om\left[1-\frac{\tilde\Pi^{\frac1\g}(-m^2)}{2\om^2}\right],\label{kdus}
\ee
where $|\tilde\Pi^{1/\g}(-m^2)|\ll\om^2$ in the perturbative regime. In fact, 
\ba
&&\frac{1}{(k_D^2+\om^2)^\g-\tilde\Pi(-m^2)}=\frac{\s(k^2_D)}{k_D^2+\tilde\om^2}\,,\\
&& \s(k^2_D)=\frac{k_D^2+\om^2-\tilde\Pi^\frac{1}{\g}(-m^2)}{(k_D^2+\om^2)^\g-\tilde\Pi(-m^2)}=\frac{k_D^2+\tilde\om^2}{(k_D^2+\om^2)^\g-(\om^2-\tilde\om^2)^\g}\,,
\ea
where the function $\s$ is finite at $k_D=\pm\rmi\tilde\om$, as one can check by applying L'H\^opital rule:
\be
\lim_{k_D^2\to-\tilde\om^2}\s(k_D^2)=\lim_{k_D^2\to-\tilde\om^2}\frac{1}{\g(k_D^2+\om^2)^{\g-1}}=\frac{1}{\g\tilde\Pi^\frac{\g-1}{\g}(-m^2)}\neq 0\,.
\ee
The fate of the branch points $k_D=\pm\rmi\om$ and of the corresponding cuts depends on the value of $\g$. If $\tilde\Pi^{1/\g}(-m^2)\in\mathbb{C}$ is complex-valued with non-vanishing imaginary part, the theory contains unstable modes and unitarity is violated. This can be avoided if $\tilde\Pi(-m^2)>0$, which from \Eq{tip3} leads to
\ba
&&\frac{\Gamma\left(2\g-\frac{D}{2}\right)}{\Gamma(2\g)}{}_2F_1\left(\g,\,2\g-\frac{D}{2};\,\g+\frac12;\,\frac{1}{4}\right)>0\qquad\Longleftrightarrow\nonumber\\
&&\g> \frac{D}{4}\,,\qquad \frac{D-4}{4}<\g<\frac{D-2}{4}\,,\qquad\dots\,,\label{1loopg}
\ea
where we wrote only the first two allowed intervals of $\g$. In particular, in four dimensions unitarity may be preserved if $\g>1$ or $0<\g<1/2$ or for other negative intervals we do not show here. For the ranges \Eq{1loopg}, since $\tilde\Pi^{1/\g}(-m^2)>0$ then $\tilde\om<\om$ and the propagator has two branch cuts and two simple poles on the imaginary axis closer to the origin than the branch points. The integration contour $\Gamma_{\rm E}$ in figure \ref{fig5} is modified by adding the pole $k_D=\rmi\tilde\om$ below $k_D=\rmi\om$ and making a circle $C_\ve^{\tilde\om}$ around it. 

To check one-loop unitarity for the massive case with cubic potential, we recalculate the contour integral with the corrected propagator. Now the presence of the simple pole adds a contribution from the residue of the integrand $\rme^{\rmi k_Dr_{ij}}/[(k_D^2+\om^2)^\g-\tilde\Pi]$ calculated at $\tilde\om$. Schematically,
\be\nonumber
\int_{C_\ve^{\tilde\om}}=\int_{\G_{\rm E}}=\int_\mathbb{R}+\int_{C_\ve}+\int_{\rm cut}+\int_{\rm arcs}=\int_\mathbb{R}+\int_{C_\ve}+\int_{\rm cut},
\ee
so that
\be\label{intint}
\int_\mathbb{R}=\int_{C_\ve^{\tilde\om}}-\int_{C_\ve}-\int_{\rm cut}
\ee
With the understanding that $\tilde\Pi=\tilde\Pi(-m^2)$ and using a change of variable similar to the one in \Eq{circle}, with $k_D-\rmi\tilde\om=\ve\exp(\rmi\theta)$, where $0<\theta<2\pi$, the pole contribution is
\ba
\fl \int_{C_\ve^{\tilde\om}}\frac{\rmd k_D}{2\pi}\,\frac{\rme^{\rmi k_Dr_{ij}}\s(k^2_D)}{k_D^2+\tilde\om^2} &=& \rmi\ve\int_{0}^{2\pi}\frac{\rmd \theta\,\rme^{\rmi\theta}}{2\pi}\,\frac{\rme^{\rmi (\ve\rme^{\rmi\theta}+\rmi\tilde\om)r_{ij}}\s(\rmi\tilde\om+\ve\rme^{\rmi\theta})}{\ve\rme^{\rmi\theta}(2\rmi\tilde\om+\ve\rme^{\rmi\theta})}\nonumber\\
\fl &=& -\frac{\rme^{-\tilde\om r_{ij}}}{2\pi\rmi(2\rmi\tilde\om)}\frac{1}{\g(\om^2-\tilde\om^2)^{\g-1}}\int_{0}^{2\pi}\rmd \theta+O(\ve)\nonumber\\
\fl &=& \frac{\rme^{-\tilde\om r_{ij}}}{2\tilde\om\g(\om^2-\tilde\om^2)^{\g-1}}+O(\ve)>0\,,
\ea
which is finite and positive. The contour integral around the branch point is
\ba
\fl \int_{C_\ve}\frac{\rmd k_D}{2\pi}\,\frac{\rme^{\rmi k_Dr_{ij}}}{(k_D^2+\om^2)^\g-(\om^2-\tilde\om^2)^\g} &=&\rmi\ve\int_{\frac{\pi}{2}}^{-\frac{3\pi}{2}}\frac{\rmd \theta\,\rme^{\rmi\theta}}{2\pi}\,\frac{\rme^{\rmi (\ve\rme^{\rmi\theta}+\rmi\om)r_{ij}}}{[\ve\rme^{\rmi\theta}(2\rmi\om+\ve\rme^{\rmi\theta})]^\g-(\om^2-\tilde\om^2)^\g}\nonumber\\
\fl &=& \rmi\ve\int_{\frac{\pi}{2}}^{-\frac{3\pi}{2}}\frac{\rmd \theta\,\rme^{\rmi\theta}}{2\pi}\,\frac{\rme^{-\om r_{ij}}}{\ve^\g\rme^{\rmi\theta\g}(2\rmi\om)^\g-(\om^2-\tilde\om^2)^\g}+\dots\nonumber\,,
\ea
which vanishes for any $\g\neq 0$. In fact, if $\g>0$ then $\int_{C_\ve}=O(\ve)$, while if $\g<0$ then $\int_{C_\ve}=O(\ve^{1-\g})$, which tend to zero as $\ve\to 0$. Finally, the contribution of the cut in \Eq{intint} replacing the free-field case \Eq{finap} reads
\ba
\fl-\int_{\rm cut}\frac{\rmd k_D}{2\pi}\,\frac{\rme^{\rmi k_Dr_{ij}}}{(k_D^2+\om^2)^\g-\tilde\Pi}
&=& \int_{\infty}^{\om}\frac{\rmd\rho}{2\pi\rmi}\,\frac{\rme^{-\rho r_{ij}}}{(\rme^{\rmi\pi}\rho^2+\om^2)^\g-\tilde\Pi}+\int_{\om}^{\infty}\frac{\rmd \rho'}{2\pi\rmi}\,\frac{\rme^{-\rho'r_{ij}}}{(\rme^{-\rmi\pi}\rho^{'2}+\om^2)^\g-\tilde\Pi}\nonumber\\
\fl&=& \int_{\om}^{\infty}\frac{\rmd \rho}{2\pi\rmi}\left[\frac{\rme^{-\rho r_{ij}}\rme^{\rmi\pi\g}}{(\rho^2-\om^2)^\g-\rme^{\rmi\pi\g}\tilde\Pi}-\frac{\rme^{-\rho r_{ij}}\rme^{-\rmi\pi\g}}{(\rho^2-\om^2)^\g-\rme^{-\rmi\pi\g}\tilde\Pi}\right]\nonumber\\
\fl&=& \int_{\om}^{\infty}\frac{\rmd \rho}{2\pi\rmi}\frac{2\rmi\sin(\pi\g)\rme^{-\rho r_{ij}}(\rho^2-\om^2)^\g}{(\rho^2-\om^2)^{2\g}-2\cos(\pi\g)(\rho^2-\om^2)^\g\tilde\Pi+\tilde\Pi^2}\nonumber\\
\fl&=& \frac{\sin(\pi\g)}{\pi}\int_{\om}^{\infty}\rmd \rho\,\frac{\rme^{-\rho r_{ij}}(\rho^2-\om^2)^\g}{(\rho^2-\om^2)^{2\g}-2\cos(\pi\g)(\rho^2-\om^2)^\g\tilde\Pi+\tilde\Pi^2}.\nonumber\\\label{finap1l}
\ea
Since the integrand is always positive for $\tilde\Pi\ll\om^2$, this integral and also the final result $I_\mathbb{R}^{ij}$ are positive semi-definite provided $\sin(\pi\g)\geq 0$, as in the non-interacting case. Therefore, we conclude that the conditions on $\g$ for one-loop unitarity are the same as the tree-level ones, for any real value of the mass $m$. This is consistent with the one-loop calculation of section \ref{dypro} using the spectral function.

\end{appendices}


\end{document}